\documentstyle[aps,prd,psfig]{revtex}

\makeatletter
\newbox\slashbox \setbox\slashbox=\hbox{\large$/$}
\def\pslash#1{\setbox\@tempboxa=\hbox{$#1$}
  \@tempdima=0.5\wd\slashbox \advance\@tempdima 0.5\wd\@tempboxa
  \copy\slashbox \kern-\@tempdima \box\@tempboxa}
\def\slash{\protect\pslash}
\makeatother

\begin{document}
\draft

\title{Statistical analysis and the equivalent of a Thouless energy in lattice
       QCD Dirac spectra}
\author{T. Guhr$^1$, J.-Z. Ma$^1$, S. Meyer$^2$ and  T. Wilke$^1$}

\address{$^1$Max Planck Institut f\"ur Kernphysik, Postfach 103980,
D-69029 Heidelberg, Germany\\
$^2$Fachbereich Physik, Universit\"at Kaiserslautern,
Theoretische Physik, D-67663 Kaiserslautern, Germany}

\date{\today}

\maketitle

\begin{abstract}
  Random Matrix Theory (RMT) is a powerful statistical tool to model
  spectral fluctuations. This approach has also found fruitful
  application in Quantum Chromodynamics (QCD).  Importantly, RMT
  provides very efficient means to separate different scales in the
  spectral fluctuations.  We try to identify the equivalent of a
  Thouless energy in complete spectra of the QCD Dirac operator for
  staggered fermions from SU(2) lattice gauge theory for different
  lattice size and gauge couplings.  We focus on the bulk of the
  spectrum.  In disordered systems, the Thouless energy sets the
  universal scale for which RMT applies.  This relates to recent
  theoretical studies which suggest a strong analogy between QCD and
  disordered systems.  The wealth of data allows us to analyze several
  statistical measures in the bulk of the spectrum with high quality.
  We find deviations which allows us to give an estimate for this
  universal scale.  Other deviations than these are seen whose
  possible origin is discussed.  Moreover, we work out higher order
  correlators as well, in particular three--point correlation
  functions.
\end{abstract}

\pacs{PACS numbers: 11.30.Rd, 05.50.+q, 64.60.Cn, 12.38.Gc}

\section{Introduction}
\label{sec1}
It is now well established that Random Matrix Theory (RMT) accurately
models spectral fluctuations in an abundant variety of different
systems, such as chaotic, disordered and many--body systems, see the
review in Ref.~\cite{guhr97}. In recent years, RMT has in addition
been successfully introduced into the study of certain aspects of
Quantum Chromodynamics (QCD). The interest focuses on the spectral
properties of the Euclidean Dirac operator.  The eigenvalue equation
under consideration reads
\begin{equation}
i\slash D[A]\psi_k=\lambda_k[A]\psi_k \ ,
\label{eq1}
\end{equation}
where $i\slash D[A]=i\slash\partial+g\slash A^at^a$ is the massless
Euclidean Dirac operator. The coupling constant is denoted by $g$ and
the $t^a$ are the generators of the gauge group. The distribution of
the color gauge fields $\slash A^a$ is given by the Euclidean QCD
partition function. As these gauge fields vary over the ensemble of
gauge field configurations, the eigenvalues fluctuate about their
average positions. The average spectral density is defined as
\begin{equation}
\rho(\lambda)=\left\langle\sum\nolimits_k\delta(\lambda-\lambda_k[A])
   \right\rangle_A \;.
\label{eq2}
\end{equation}
The average has to be performed over all gauge field configurations.

In contrast to most other systems, however, there are two
different regimes in QCD spectra which can be addressed in an RMT
approach, the microscopic region and the bulk region. 
Since the Dirac operator only couples states of opposite chirality,
the eigenvalues are pairwise positive and negative. This is the reason
why two types of spectral fluctuations can be distinguished, namely
spectral fluctuations in the microscopic limit near zero virtuality,
$\lambda=0$, and in the bulk of the spectrum. 
 
Concerning the microscopic region, chiral Random Matrix Theory (chRMT)
\cite{shuryak93} incorporates the global symmetry properties, in
particular chiral symmetry, of $i\slash D$. It predicts level
repulsion between positive and negative eigenvalues which results in a
distinct behavior of the eigenvalue density and correlations near the
origin.  It is possible to calculate spectral correlators analytically
in the microscopic limit \cite{verbaar93,forr93,nagao95}, and to
compare the predictions of chRMT with complete spectra of the lattice
QCD Dirac operator on reasonably large lattices.  Indeed, remarkable
agreement is found~\cite{bitsch97a,bitsch97b,ma97,verbaar96,verbaar97}
at the edge of the spectrum.

Sufficiently far away from the origin, however, the repulsion of
negative and positive eigenvalues should become unimportant.
Therefore the chiral structure of the theory is not expected to be of
relevance in the bulk of the spectrum.  This is the region we will
address on in this work. By comparing with lattice data, it has
already been shown that conventional RMT properly models these
fluctuations in the bulk \cite{halasz95,halasz97,verbaar97}.  It is
important to go beyond these statistical analyses made so far in order
to see to what scales RMT does apply. The identification of such
scales gives a fundamental insight into a system. Investigations of
this type have been performed in great detail in disordered systems
and in many--body systems.  There, the Thouless energy $E_c$ or the
spreading width $\Gamma$ determine the scale $E_c/D$ where $D$ is the
level spacing, in which the fluctuation are of RMT type
\cite{thou75,thou77}.  Beyond this scale, deviations from the RMT
behavior occur, see Ref.\cite{guhr97} for a detailed discussion and
further references.

Recently, theoretical studies \cite{stern98,osborn98,janik98}
established a link between disordered systems and QCD. The range of
validity of RMT, $\lambda_{\rm RMT}$, was introduced as an equivalent
of a Thouless energy $E_c$. A scaling $\lambda_{\rm RMT}/D \propto
\sqrt{V}$ was proposed where $V$ is the four--volume of the system.
Indeed, such a scaling behavior was found very recently in the
microscopic region \cite{bitsch98} for deviation from RMT behavior. As
argued in \cite{osborn98} a corresponding effect should also be seen
in the bulk of the spectrum. This is what we will investigate.

The identification of this scale $\lambda_{\rm RMT}/D$ in QCD spectra
could lead to an improved understanding of certain features of QCD and
allows us to separate the stochastic noise of the short range
fluctuations from the true dynamics of the QCD vacuum. Eventually, it
could be possible to set up effective models or simplify the presently
used simulation algorithms in lattice gauge theories.  In this work,
we search for such a universal scale $\lambda_{\rm RMT}/D$ in the bulk
region by analyzing lattice data.  In contrast to the microscopic
region, the bulk of the spectrum is expected to have a translation
invariant analogue of the Thouless energy.  We emphasize that our
analysis is self--consistent. Advantageously, it does not depend on
any model that aims at an explanation for the occurrence of this
universal scale.

The high amount and quality of the data sets which exceed the
existing ones by far enable us to considerably extend the energy
range for our analysis.  Moreover, the wealth of data makes it
possible to directly address bare correlation functions which cannot
be analyzed in most systems. Furthermore, in doing so we discuss some
technical aspects, which are of general interest for the investigation
of spectral fluctuations.

This paper is organized as follows: In Sec.~\ref{sec2} the data under
investigation is presented.  A detailed analysis of the statistical
properties is given in Sec.~\ref{sec3}. This includes the introduction
of the numerical unfolding approaches, a statistical analysis of the
nearest neighbor spacing distribution, two--point spectral
correlations and higher order spectral correlations. Deviations from
the RMT predictions are found and interpreted. Summary and discussion
are given in Sec.~\ref{sec4}.

\section{Complete Dirac spectra in SU(2) gauge theory}
\label{sec2}

The computation of large ensembles of complete spectra of the
Euclidean Dirac operator for staggered fermions in SU(2) gauge theory
has recently been performed Ref.\cite{MBB98} expanding the numerical
work of \cite{Kalk95}. In lattice gauge theory simulations one
generates a sequence of gauge field configurations distributed
according to the Boltzmann weight. On each of the gauge field
configurations the eigenvalue equation (\ref{eq1}) is solved
numerically on the lattice and a distinct partition of eigenvalues is
obtained.  The lattices have the size $V=L^4$ where $L$ denotes the
length of the Euclidean box with a lattice spacing $a$.  Parameters
and statistics of the simulation are summarized in Table\ref{tab1}.
The choice of SU(2) as the gauge group implies that every eigenvalue
of $i\slash D$ is twofold degenerate due to a global charge
conjugation symmetry. The random--matrix ensemble for this situation
has symplectic symmetry and is referred to as chiral Gaussian
Symplectic Ensemble (chGSE) \cite{verbaar94,guhr97}.  In addition, the
squared Dirac operator $-\slash D^2$ couples only even to even and odd
to odd lattice sites, respectively.  Thus, $-\slash D^2$ has $V/2$
distinct eigenvalues.  We use the Cullum--Willoughby version of the
Lanczos algorithm \cite{Stoe93} to compute the complete eigenvalue
spectrum of the sparse hermitian matrix $-\slash D^2$ in order to
avoid numerical uncertainties for the low-lying eigenvalues.  There
exists an analytical sum rule, ${\rm tr}(-\slash D^2) = 4V$, for the
distinct eigenvalues of $-\slash D^2$ \cite{Kalk95}.  We have checked
that this sum rule is satisfied by our data, the largest relative
deviation was about $10^{-8}$.

Examples of the spectra are shown in Fig.~\ref{stair}, where the
average level density and the integrated average level density, see
Eq.~(\ref{staircase}), for a $16^4$, i.e.~$L=16a$, lattice are shown.
It should be pointed out that due to the $V/2=32768$ distinct
eigenvalues of each configuration there are millions of eigenvalues at
our disposal. We used two different values of the gauge coupling
$\beta=4/g^2$ where the weak coupling regime of SU(2) sets in and
where most of the scaling test have been performed so far.  Finally,
the chiral condensate was obtained by fitting the spectral density and
extracting $\rho(0)$.  Our findings \cite{MBB98} are in rough
agreement with the values obtained by Hands and Teper \cite {Hands90}
for the same simulation parameters in SU(2) but only the 20 smallest
eigenvalues have been computed by these authors.

\section{Data analysis}
\label{sec3}

In this section we give a detailed analysis of the data introduced in
the previous section in the bulk of the spectrum. We start with a
description of the numerical unfolding approaches and their properties
in Sec.~\ref{sec31}. After a short discussion of the nearest neighbor
distribution in Sec.~\ref{sec32}, we present data for spectral
two--point correlations at large scales in Sec.~\ref{sec33}.  From this
we identify the equivalent of a Thouless energy.  Furthermore we
discuss higher order correlators, in particular three--point
correlations in Sec.~\ref{sec34}. A qualitative explanation of
deviations from RMT predictions which are not due to the Thouless
energy is given in Sec.~\ref{sec35}.

\subsection{Unfolding}
\label{sec31}

As RMT is capable of making predictions for the fluctuations on the
scale of the mean level spacing, one has to remove the influence of
the level density by unfolding the spectra.  The cumulative spectral
function
\begin{equation}
N(\lambda)=
        \int_{-\infty}^\lambda d\lambda^\prime
        \sum_{i=1}^{V/2}\delta(\lambda^\prime-\lambda_i)\;,
\label{staircase}
\end{equation}
is the number of levels below or at the energy $\lambda$. It is
frequently referred to as staircase function.  It can be separated
into an average part $N_{\rm ave}(\lambda)$, whose derivative is the
level density, and a fluctuating part $N_{\rm fluc}(\lambda)$,
\begin{equation}
  N(\lambda)=N_{\rm ave}(\lambda)+N_{\rm fluc}(\lambda)\;.
\end{equation}
The average part is determined by gross features
of the system and has to be removed.  The fluctuating part is in
all relevant systems of order ${\cal O}(1)$ and contains the
correlations to be analyzed. After extraction of the average part
$N_{\rm ave}(\lambda)$, it is unfolded from the spectra by the
introduction of a dimensionless energy variable
\begin{equation}
\xi_i=N_{\rm ave}(\lambda_i) \ .
\end{equation}
In this variable, the spectra have mean level spacing unity
everywhere,
\begin{equation}
1/\rho_{\rm ave}(\xi)\equiv 1,
\label{lsone}
\end{equation}
where $\rho_{\rm ave}(\xi)=dN_{\rm ave}(\xi)/d\xi$. However, the
extraction of $N_{\rm ave}(\lambda)$ from the data is non--trivial in
our case because little is known analytically about the level density
of QCD spectra, particularly in lattice calculations. We thus have to
resort to phenomenological unfolding procedures.  Faulty unfolding
leads to wrong results, especially on such large energy scales that we
are interested in. In the subsequent Secs.~\ref{sec311},
\ref{sec312} and \ref{sec313}, we discuss three different procedures
used here, ensemble unfolding, configuration unfolding and windowing,
respectively.

\subsubsection{Ensemble unfolding}
\label{sec311}
In RMT one deals with an ensemble of matrices, where the matrix
elements of each member are chosen randomly. Spectral observables
predicted by RMT are calculated as an average over the ensemble.  This
ensemble average is denoted by a bar $\overline{(\ldots)}$.  But
observables can also be calculated as spectral average, i.e.  one
performs a running average over overlapping intervals
$[\alpha,\alpha+L]$ of length $L$ in the spectrum of one member.  In
order to distinguish it from ensemble average, we denote spectral
averaging by angular brackets $\langle\ldots\rangle$. In the limit of
large matrix dimension both averages are equivalent \cite{guhr97}.
This property is called ergodicity.

\begin{figure}
\centerline{\psfig{figure=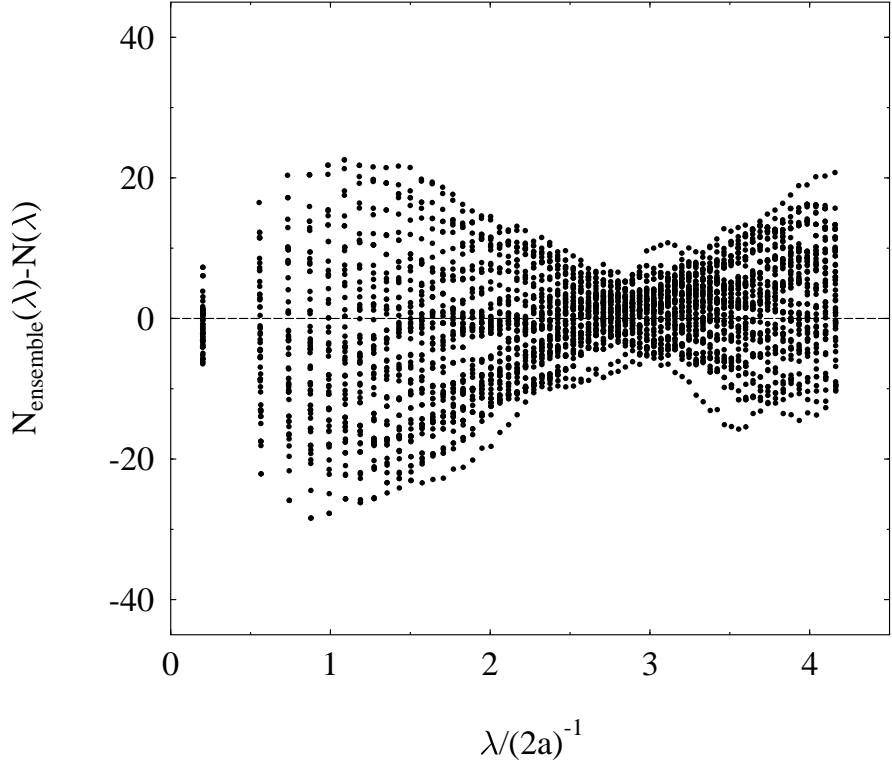,width=16cm,angle=0}}
\caption{ Difference between the integrated level density
  $\overline{N}(\lambda)$ averaged over all 921 configuration
  ($\beta=2.4$) and real data. Each dot represents the value of
  $\overline{N}(\lambda_{i,j})-N(\lambda_{i,j})$. Index $i$ enumerates
  the eigenvalues, $i=1,\ldots,32768$, and $j$ is the configuration
  number, $j=1,\ldots,50$. The $50$ plotted configuration where chosen
  arbitrarily. Only every 500th eigenvalue is shown.
\label{ensemble}}
\end{figure}

In most experiments, one measures one - preferably long - spectrum.
Thus observables are usually calculated from spectral average. One
uses the theoretical concept of ergodicity to compare the RMT
predictions with the experimental results.  In our case, however, the
data consists of configurations, i.e. forms an ensemble. Hence,
questions related to ergodicity arise not only for the calculation of
observables, but also in the determination of the staircase function,
i.e.~in the unfolding procedure.  We have in principle two very
different ways of unfolding our data: first, ensemble unfolding,
$N_{\rm ave}(\lambda)=\overline{N}(\lambda)$, i.e.~we determine the
smooth part of the staircase function by averaging over the ensemble,
and second, configuration unfolding, $N_{\rm ave}(\lambda)=\langle
N(\lambda)\rangle$, i.e.~we determine the smooth part of the staircase
function for every configuration separately. The results differ
considerably, $\overline{N}(\lambda)\neq\langle N(\lambda)\rangle$,
for most of the configurations.  The ensemble averaged staircase
$\overline{N}(\lambda)$ for the lattice QCD Dirac operator is shown in
Fig.~\ref{stair}.  We find $\overline{N}(\lambda)$ by dividing the
energy range in $m$ bins with width $\Delta\lambda$ and average the
density $\rho(\lambda,\lambda+\Delta\lambda)$ for each bin over all
configurations. We then calculate the staircase function as
$\overline{N}(\lambda)=\sum_{i=1}^m\rho(\lambda_i,
\lambda_i+\Delta\lambda)\Delta\lambda$, with $\lambda_m=\lambda$.  In
Fig.~\ref{ensemble}, the difference between the ensemble averaged
staircase function and the configuration wise averaged ones, for
$V=16^4$ and $\beta=2.4$, for 50 arbitrarily chosen configurations is
plotted.  Each data point represents the difference
$\overline{N}(\lambda_{i,j})-\langle N(\lambda_{i,j})\rangle$, where
$i$ enumerates the eigenvalues, $i=1,\ldots,32768$ and $j$ is the
configuration number $j=1,\ldots,50$. We plot only every 500th
eigenvalue.  There are deviations of about
$\overline{N}(\lambda)-\langle N(\lambda)\rangle={\cal O}(10^1)$ in
certain energy ranges.

If the spectra are unfolded using the ensemble averaged staircase
function $\overline{N}(\lambda)$, observables should then also be
calculated as an ensemble average for a fixed value of $\lambda$. But
we checked that our results do not depend on $\lambda$ in a wide range
of the bulk.  This property is called translational invariance. It is
actually not present in the microscopic region, where it is destroyed
by point wise symmetry between positive and negative eigenvalues
\cite{bitsch98}. Translational invariance in the bulk allows us to
calculate observables from running average over overlapping intervals
for each configuration. We choose an overlap of $90\%$ for two
consecutive intervals. Then we average over all configurations. This
improves the statistics of the result considerably.

\subsubsection{Configuration unfolding}
\label{sec312}
We now unfold each configuration separately.  Observables are then
calculated for each configuration by running spectral average.
Thereafter we average over the ensemble. The basic characteristics are
already obtained for one single configuration, though the statistics
is considerably improved by ensemble averaging.  This is in the same
spirit as it was done in spectra of nuclei \cite{bohigas83} and
complex atoms \cite{rosen60}.  These spectra were unfolded for each
nuclei or atom separately.  Then observables were calculated as
described above, i.e. first taking the spectral and then the ensemble
average. In this case the ensembles consist of nuclei or atoms of
different types.

Configuration unfolding is, in contrast to ensemble unfolding, not a
unique procedure. One has to find either fits to the average staircase
function or to smooth it in some way. A priori, there is no criterion
whether the numerical estimated $N_{\rm ave}(\lambda)$ is close to the
real one or not.  Thus, we use three different approaches and
carefully compare them with one another to eliminate as many sources
for mistakes as possible and to obtain consistent results.

First, we fit $N(\lambda)$ to a polynomial of degree $n$,
\begin{equation}
\langle N(\lambda)\rangle
 =N_{\rm poly}(\lambda)=\sum_{j=0}^n\:a_j\,\lambda^j\;,
\label{poly}
\end{equation}
where $n$ is a small integer, $n=2,\ldots,5$.  This approach is
motivated by the fact that almost all physical systems are known to
have a level density which is as smooth as a polynomial. In our case
this ansatz is supported by pertubative calculations. Strong coupling
expansions for SU(2) with staggered fermions have been performed
\cite{kogut83} and, furthermore, $1/N_c$ expansion of the QCD level
density \cite{smilga93}, both suggesting a smooth level density. The
former gives a semi--circle whereas the latter explicitly predicts a
polynomial increase.

Second, we use the Gaussian method which was originally developed by
Strutinsky \cite{haake}.  One replaces the $\delta$--functions in
Eq.~(\ref{staircase}) by Gaussian functions with a width $\Delta$
which yields a smoothed staircase
\begin{equation}
N_\Delta(\lambda)=\int_{-\infty}^\lambda d\lambda'\:
        \sum_{i={\rm min}}^{{\rm max}}\:\frac{1}{\sqrt{\pi}\Delta}\,
        e^{-\frac{1}{\Delta^2}\,(\lambda'-\lambda_i)^2} .
\label{gauss}
\end{equation}
The summation runs from the smallest eigenvalue $\lambda_{\rm min}$ to
the largest eigenvalue $\lambda_{\rm max}$ in the interval under
consideration.  The limit $\Delta\to0$ restores $\delta$--distributed
eigenvalues, whereas the fluctuations are smeared out for finite
$\Delta$.  The optimal parameter $\Delta_{\rm opt}$ is found by a
$\chi^2$--fit of $N_\Delta(\lambda)$ to $N(\lambda)$. Then we identify
\begin{equation}
\langle N(\lambda)\rangle=N_{\Delta_{\rm opt}}(\lambda) \ .
\label{Ngauss}
\end{equation}

Third, we perform a local unfolding by calculating the unfolded
eigenvalues $\xi_i$ directly with the formula
\begin{equation}
\xi_{i+1}-\xi_i=\frac{\lambda_{i+1}-\lambda_i}{D_i}\;,
\end{equation}
with local mean level spacing
\begin{equation}
D_i=\frac{1}{2k+1}\sum_{j=i-k}^{i+k}(\lambda_{j+1}-\lambda_j)\;.
\label{mls}
\end{equation}
Here $2k$ is the number of consecutive level spacings over the running
average is performed.

\begin{figure}
\psfig{figure=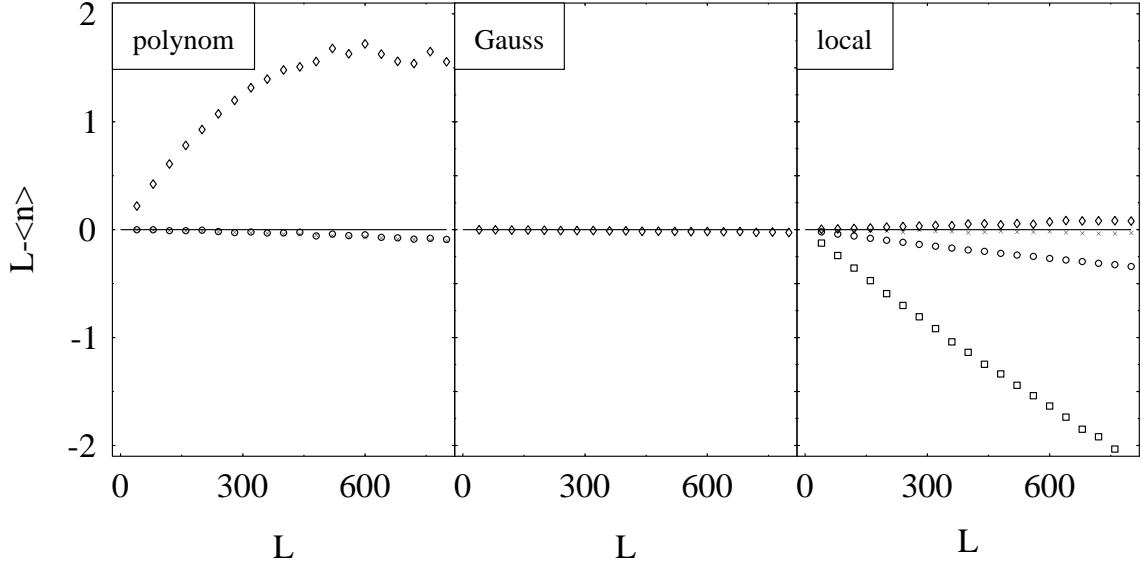,width=16cm,angle=0}
\caption{Value of the quantity
  $L-\overline{\langle n_\alpha(L)\rangle}$ for the three unfolding
  approaches on a $16^4$ lattice. From left to right the polynomial,
  Gaussian and local unfolding is shown. In the left plot diamonds are
  data for $n=2$ and the cross and circles are $n=3$ and $n=4$.  In
  the right plot the data points from top to bottom correspond to an
  averaging interval of $k=20$,100,300 and 900,
  respectively.\label{mean}}
\end{figure}

Whatever approach one decides to use, a necessary condition is that on
the unfolded scale the average number of levels in an interval of
length $L$ should equal this length. This is a very important
requirement because we are also interested in very large energy
scales.  This assures that the spectrum on the numerically constructed
dimensionless scale $\xi$ has mean level density unity.  Consider the
interval $[\alpha,\alpha+L]$ which contains $n_\alpha(L)$ eigenvalues.
Spectral average $\langle\ldots\rangle$ and ensemble average
$\overline{(\ldots)}$ have to yield
\begin{equation}
\overline{\langle n_\alpha(L)\rangle}=L\;.
\label{meannumber}
\end{equation}
In Fig.~\ref{mean}, the difference between the calculated mean number
of eigenvalues $\overline{\langle n_\alpha(L)\rangle}$ and $L$ is
plotted as a function of $L$. For the Gaussian method the difference
appears to be zero to all scales. While it is small and does only
appear at large $L$ for polynomials fits with $n\geq3$, strong
deviations from the zero line already appear at small $L$ for $n=2$.
In the case of local unfolding, the difference $L-\overline{\langle
  n_\alpha(L)\rangle}$ is positive for small $k$, i.e.  there are on
average less levels in a given interval than there should be. For
growing $k$, it becomes negative with ever stronger deviations from
the flat line. The averaging interval length $k$ for which the
difference equals zero is $k\approx100$ for $V=16^4$. For other
lattice sizes this averaging interval is slightly smaller.  We take
this as the optimal parameter for this approach. {}From Fig.~\ref{mean}
we learn, that the necessary condition (\ref{meannumber}) is fulfilled
only for the Gaussian approach, polynomial fit with $n\geq3$ and local
unfolding with $k\approx100$ for $V=16^4$, all other choices of the
parameters must be rejected.

It should be mentioned that a new artificial scale both for local and
Gaussian approach is introduced, namely the averaging interval length
$k$ and the width $\Delta$, respectively. Therefore, one should be
cautious in the interpretation of effects seen on scales $L$ larger
than current value of the corresponding parameter. In units of the
mean level spacing $D$ we find a width of the Gaussian as
$\Delta/D\approx100$ at a $16^4$ lattice. On the other hand, both
approaches have the advantage that no particular function for the
average level density has to be assumed.

We checked all our numerical unfolding approaches with the spectrum of
a very different system. We used the spectrum of quantum chaotic
billiard that was simulated in a microwave experiment. In billiards,
the Weyl formula gives an analytically expression for the mean level
density \cite{graef92}. With our phenomenological approaches, we
indeed obtained quantitatively the same results.

\subsubsection{Windowing}
\label{sec313}
Ideally, an unfolding procedure should only remove the global
variations of the spectral density, i.e. in our case the overall
behavior seen in Fig.~\ref{stair}. For reasons which will become
clearer later, it is difficult to numerically distinguish the global
variations from the local fluctuations. This is in particular the case
for data of large lattice size.  In other words, we might have removed
too much by some of the unfolding procedures, while we might have
removed too little by others. This will be discussed in great detail
below, especially in Sec.~\ref{sec35}.

One has to ensure that any deviations seen in the spectral statistics
are not due to global variations in the average density which were not
removed adequately. One way is to use different unfolding approaches
and compare the results carefully. Another way is to take only a small
window of the spectra in which the global variation of the density is
expected to be small.  Thus we choose an interval
$[\lambda,\lambda+\delta\lambda]$ and calculate the ensemble averaged
mean level spacing $D$ for it. We then rescale the eigenvalues in this
interval as
\begin{equation}
\xi_i=\lambda_i/D\quad,\quad
\lambda<\lambda_i<\lambda+\delta\lambda\;.
\end{equation}
This is done in the same manner as in \cite{ma97} where the
microscopic region was considered. However, it is not clear beforehand
that a scale in the spectra, if any, does not exceed the interval
length $\delta\lambda$. Unfolding, if done correctly, allows to make
investigations to much larger scales.

This approach is closely related to ensemble unfolding defined above.
Indeed, the results coincide, but with a less statistical significance
by only rescaling.  By using this approach, we intended to avoid any
unfolding procedures. As we will see later, there are a slight, but
still systematic variations of the spectral density within the small
window.

\subsection{Nearest Neighbor Spacing Distribution}
\label{sec32}

\begin{figure}
\psfig{figure=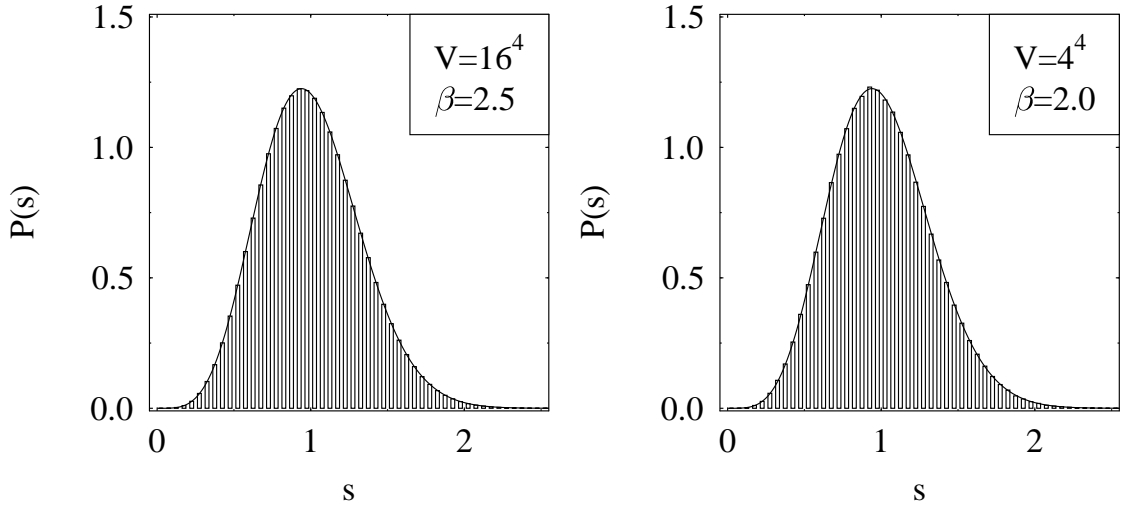,width=16cm,angle=0}
\caption{Nearest neighbor distribution, solid line is the Wigner surmise
  and the bars represent the lattice data.\label{near}}
\end{figure}

The nearest neighbor spacing distribution $P(s)$ probes the
fluctuations on short scales in the spectra. It is the probability of
finding the distance $s$ between adjacent level on the unfolded scale.
It contains all correlations of order $k\ge 2$.  In the case of
completely uncorrelated levels which is referred to as Poisson
regularity~\cite{guhr97}, it is given by $P(s)=\exp(-s)$. In the case
of GSE type correlations, Wigner surmized the shape
\begin{equation}
P(s)=\frac{262144}{729\pi^3}s^4\exp\left(-\frac{64}{9\pi}s^2\right)\;,
\end{equation}
which is very close to the exact GSE result. As shown in
Fig.~\ref{near}, the data is in perfect agreement with the prediction.
This is as well true for the large lattice, $V=16^4$ (left part), as
for the small lattice, $V=4^4$ (right part). The spacing distribution
of the intermediate lattice sizes are not distinguishable from the
both shown. We have complete agreement between theory and lattice data
for any choice of lattice size and coupling constant.

\subsection{Two--Point Spectral Correlations}
\label{sec33}

In an interval of length $L$ in units of the mean level spacing, the
mean number of eigenvalues should be equal to $L$, see
Eq.~(\ref{meannumber}). The variance of this number is defined by
\cite{Mehta}
\begin{equation}
\Sigma^2(L)=\overline{\langle(L-n_{\alpha}(L))^2\rangle} \ .
\label{sigma}
\end{equation}
Thus, an interval of length $L$ contains on average
$L\pm\sqrt{\Sigma^2(L)}$ levels.  For uncorrelated Poisson spectra
$\Sigma^2(L)=L$. RMT predicts for the number variance stronger
correlations, namely $\Sigma^2(L)\sim \log L$. 

Another important quantity is the spectral rigidity $\Delta_3(L)$,
defined as the least square deviation of the staircase function from
the straight line \cite{Mehta},
\begin{equation}
  \Delta_3(L)=\left\langle\frac{1}{L}{\rm
    min}_{A,B}\int_{\alpha}^{\alpha+L}d\xi(N(\xi)-A\xi-B)^2
           \right\rangle \ .
\end{equation}
Since it can be expressed as an integral over the number variance
\begin{equation}
\Delta_3(L)=\frac{2}{L^4}\int_0^Ldr(L^3-2L^2r+r^3)\Sigma^2(r)\;,
\label{rigidity}
\end{equation}
it is smoother than $\Sigma^2(L)$.  

The number variance can be expressed as an integral of the two--point
cluster function $Y_2(r)$ \cite{Mehta}, which depends for
translational invariant spectra only on the difference
$r=|\xi_2-\xi_1|$ between two levels at $\xi_1$ and $\xi_2$,
\begin{equation}
\Sigma^2(L)= L-2\int_{0}^{L}(L-r)Y_2(r)dr \ .
\label{mu2}
\end{equation}
The cluster function is related to the two--point correlation function
$X_2(r)$ which measures the probability density to find two levels at
a distant $r$ by $X_2(r)=1-Y_2(r)$.  In contrast to $P(s)$, these two
levels are not restricted to adjacent ones.  

\begin{figure}
\begin{minipage}{7.5cm}
  \centerline{\psfig{figure=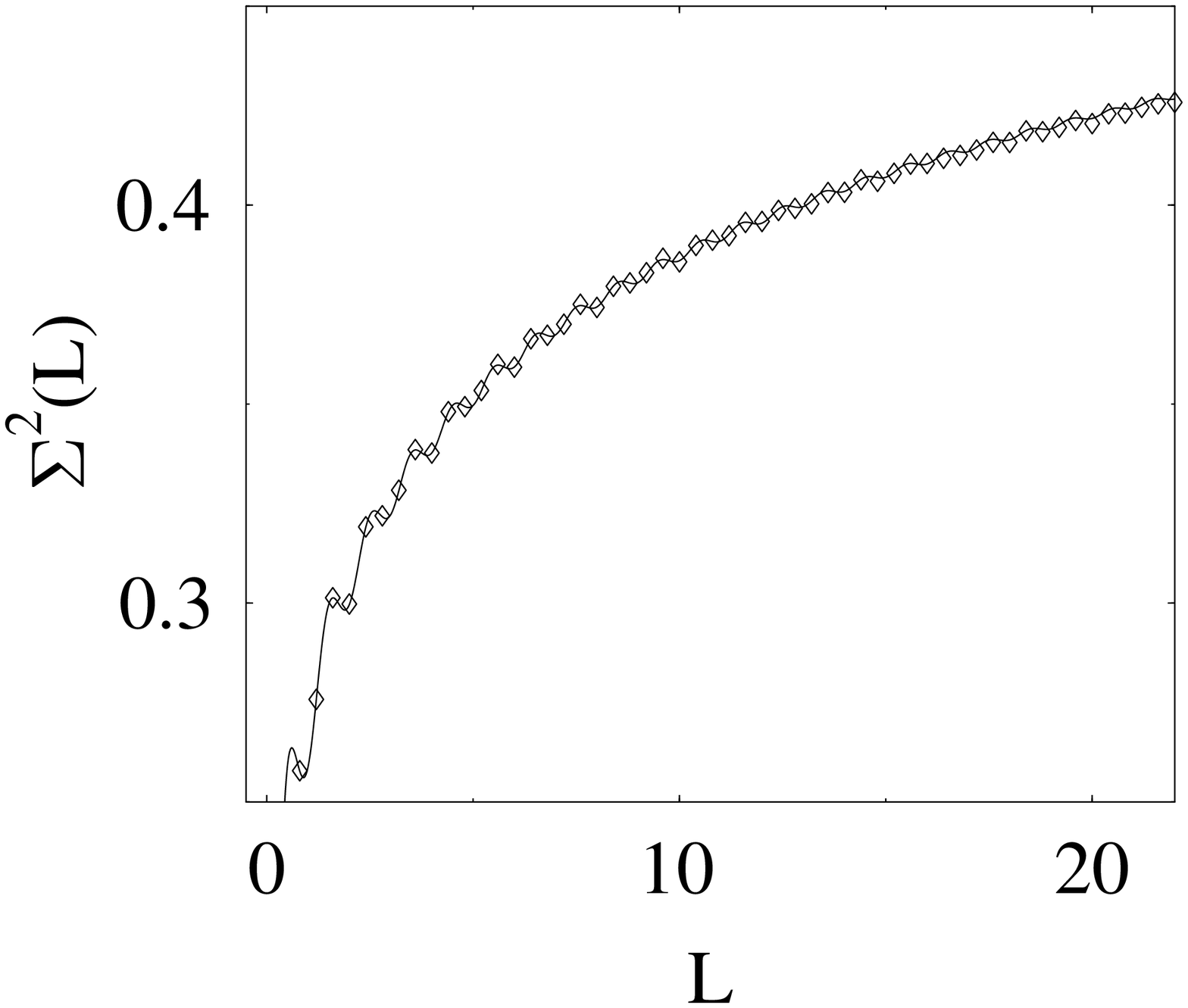,width=8cm,angle=0}}
\end{minipage}
\hfill
\begin{minipage}{7.5cm}
\centerline{\psfig{figure=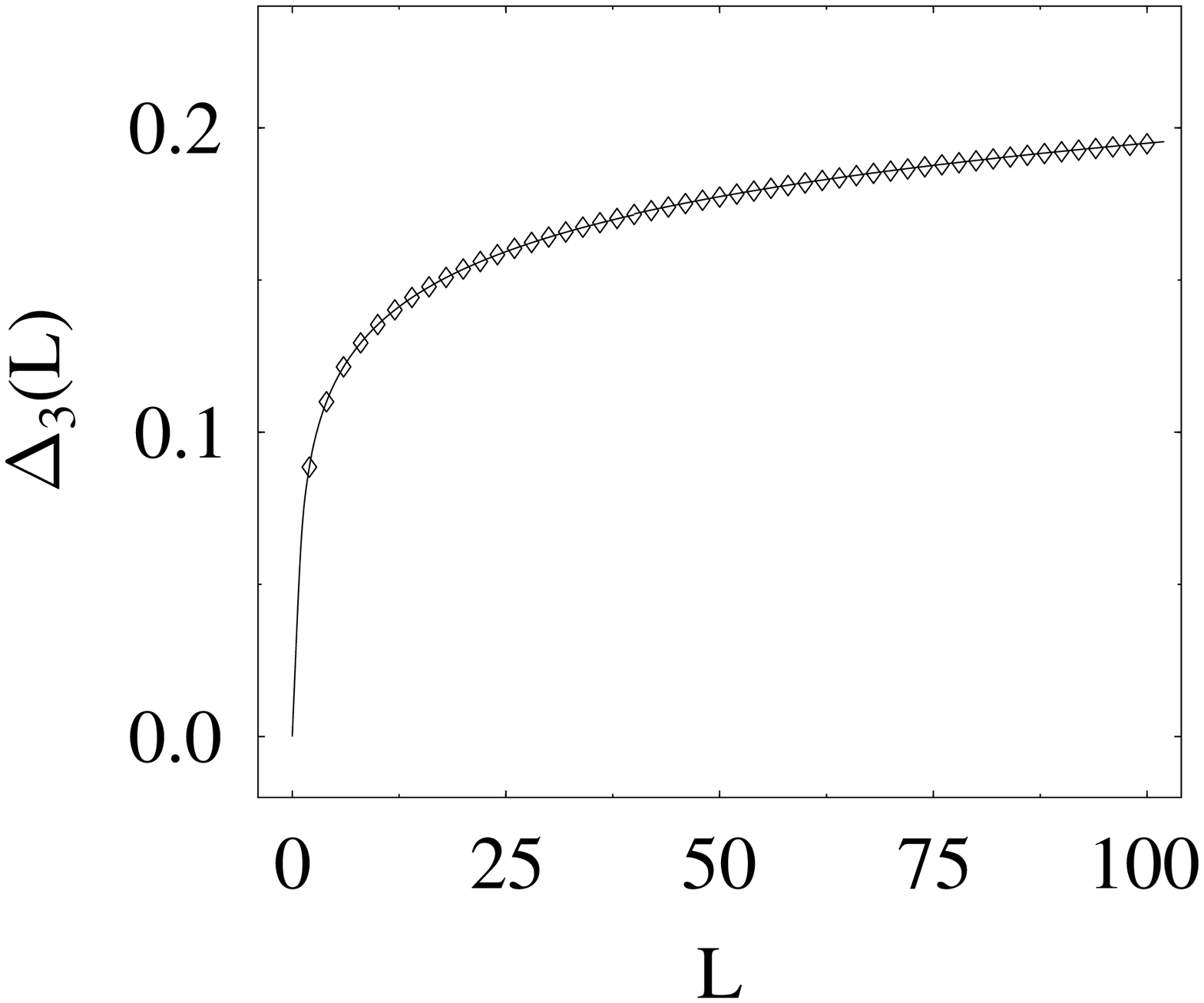,width=8cm,angle=0}}
\end{minipage}
\caption{Integrated two--point functions number variance $\Sigma^2(L)$,
  spectral rigidity $\Delta_3(L)$ for small $L$ on a $16^4$-lattice.
  The solid line represent the RMT predictions and the dots the data.
  On this scale the presented data points do not depend on unfolding.
  Note the difference in the scale of the $L$ axes between
  $\Delta_3(L)$ and $\Sigma^2(L)$.
  \label{agree1}}
\end{figure}

\begin{figure}
\begin{minipage}{7.5cm}
\centerline{\psfig{figure=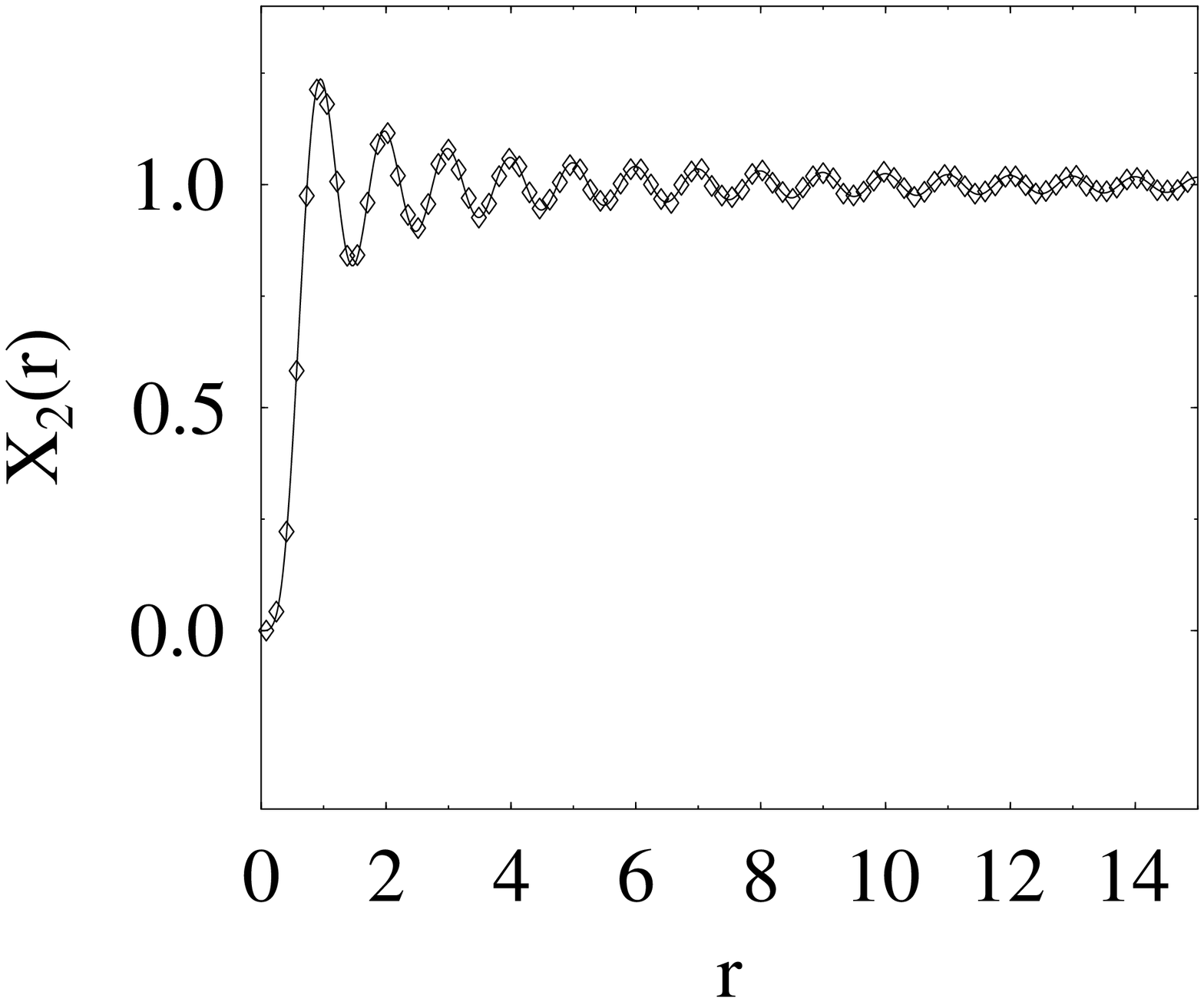,width=8cm,angle=0
}}
\end{minipage}
\begin{minipage}{7.5cm}
\centerline{\psfig{figure=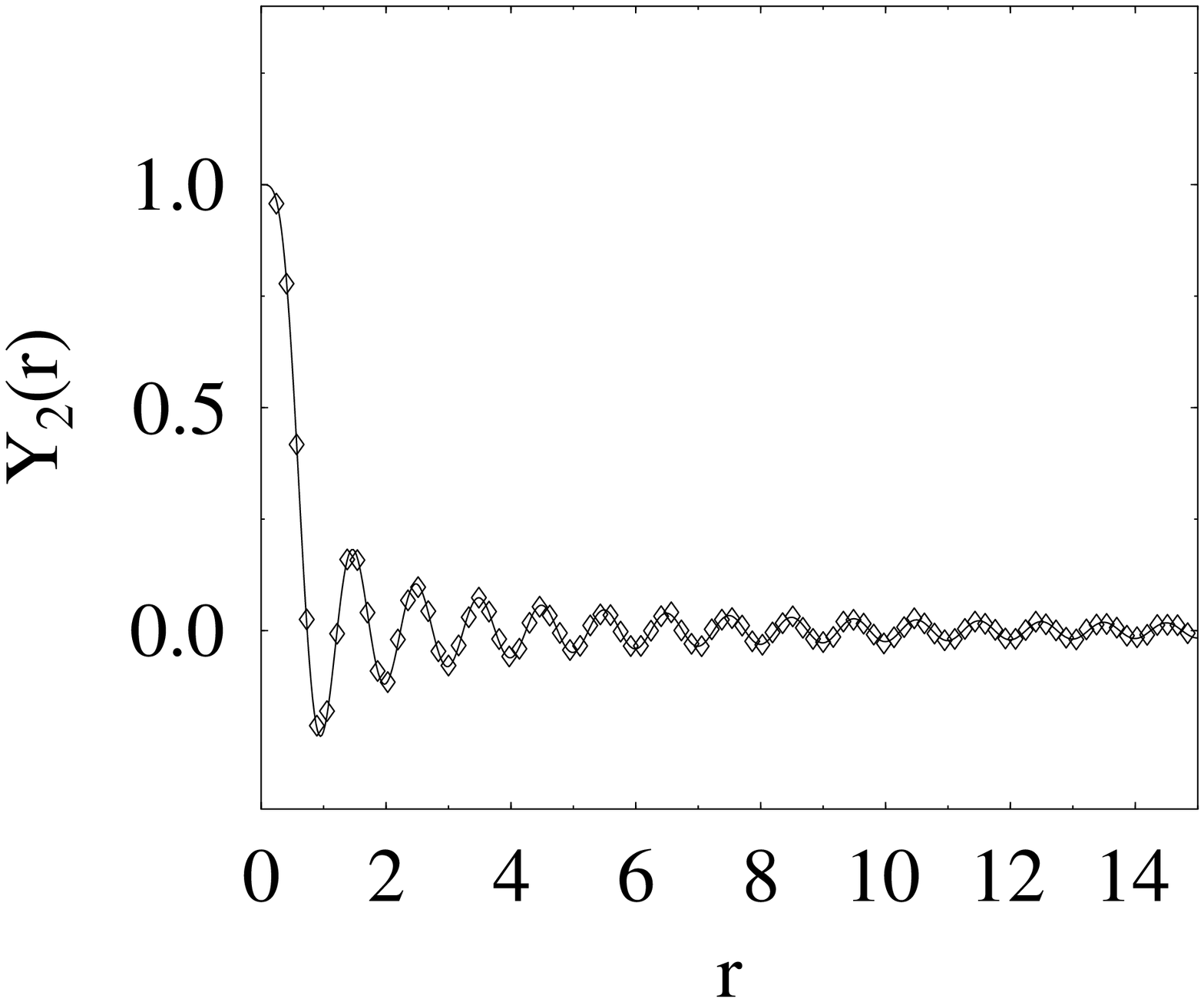,width=8cm,angle=0
}}
\end{minipage}
\caption{\label{twfig1}
  The two--point correlation function $X_2(r)$(left) and the cluster
  function $Y_2(r)$ (right) as a function of $r$, compared with the
  GSE predictions.  The result is independent of the unfolding
  approach.}
\end{figure}

In Fig.~\ref{agree1}, number variance and spectral rigidity for scales
up to $L=20$ and $L=100$, respectively, are shown. Lattice data and
RMT predictions agree remarkably well, even the oscillations in
$\Sigma_2(L)$ are accurately reproduced. Naturally, previous
analyses~\cite{halasz95} with smaller data sets have less statistical
significance.  Two--point cluster and correlation function which
usually are not accessible in data analysis are shown in
Fig.~\ref{twfig1}. Again, the agreement is impressive.

Beyond this scale there are considerable deviations of $\Delta_3(L)$
as well as of $\Sigma^2(L)$ from RMT predictions which depend on the
unfolding procedure used. We mention that on general grounds one can
show that any scales in $\Sigma^2(L)$ and $\Delta_3(L)$, say
$L^\Sigma$ and $L^\Delta$, respectively, are related by
$L^\Delta:L^\Sigma=4:1$, or so \cite{guhr97}, see Fig.~\ref{agree1}.
Thus, any deviations from RMT behavior appear at smaller $L$ in
$\Sigma^2(L)$ as compared to $\Delta_3(L)$.

\begin{figure}
\centerline{\psfig{figure=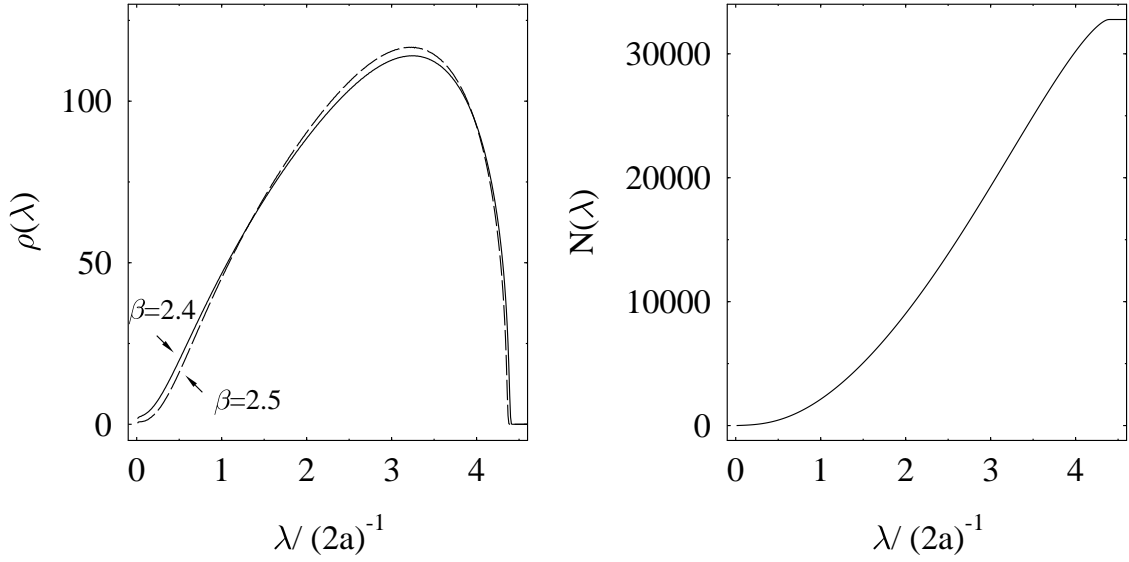,width=16cm,angle=0}}
\caption{Average level density $\bar{\rho}(\lambda)$ for $\beta=2.4$
  and $\beta=2.5$ (left plot) and integrated average level density
  $\overline{N}(\lambda)$, see Eq.(\protect\ref{staircase}), for
  $\beta=2.5$ (right plot). The eigenvalues are given in units of the
  inverse lattice spacing $(2a)^{-1}$. The bin sizes in the left plot is
  $0.01\cdot(2a)^{-1}$.
\label{stair}}
\end{figure}

\begin{figure}
\begin{minipage}{7.5cm}
  \centerline{\psfig{figure=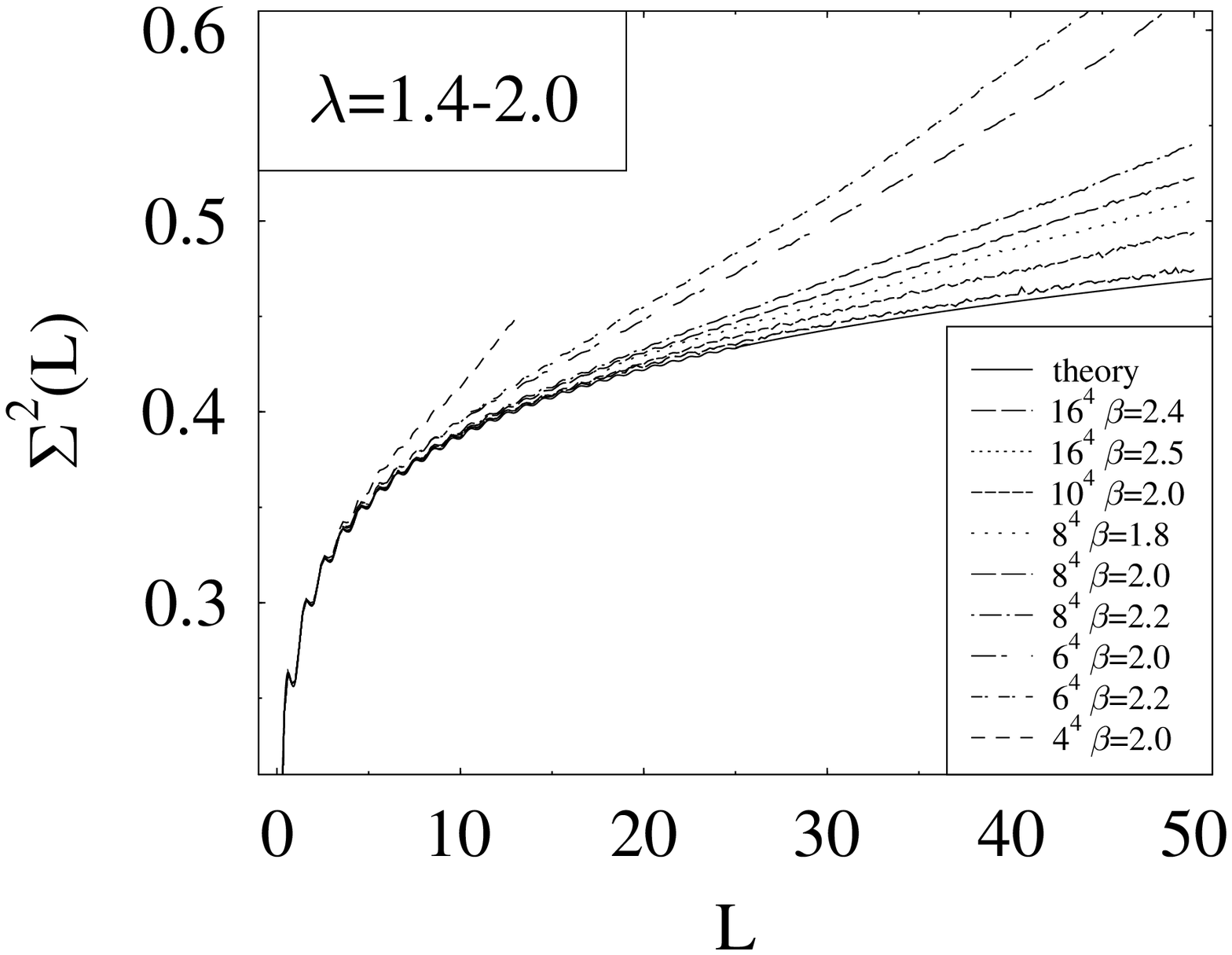,width=8cm,angle=0}}
\end{minipage}
\hfill
\begin{minipage}{7.5cm}
\centerline{\psfig{figure=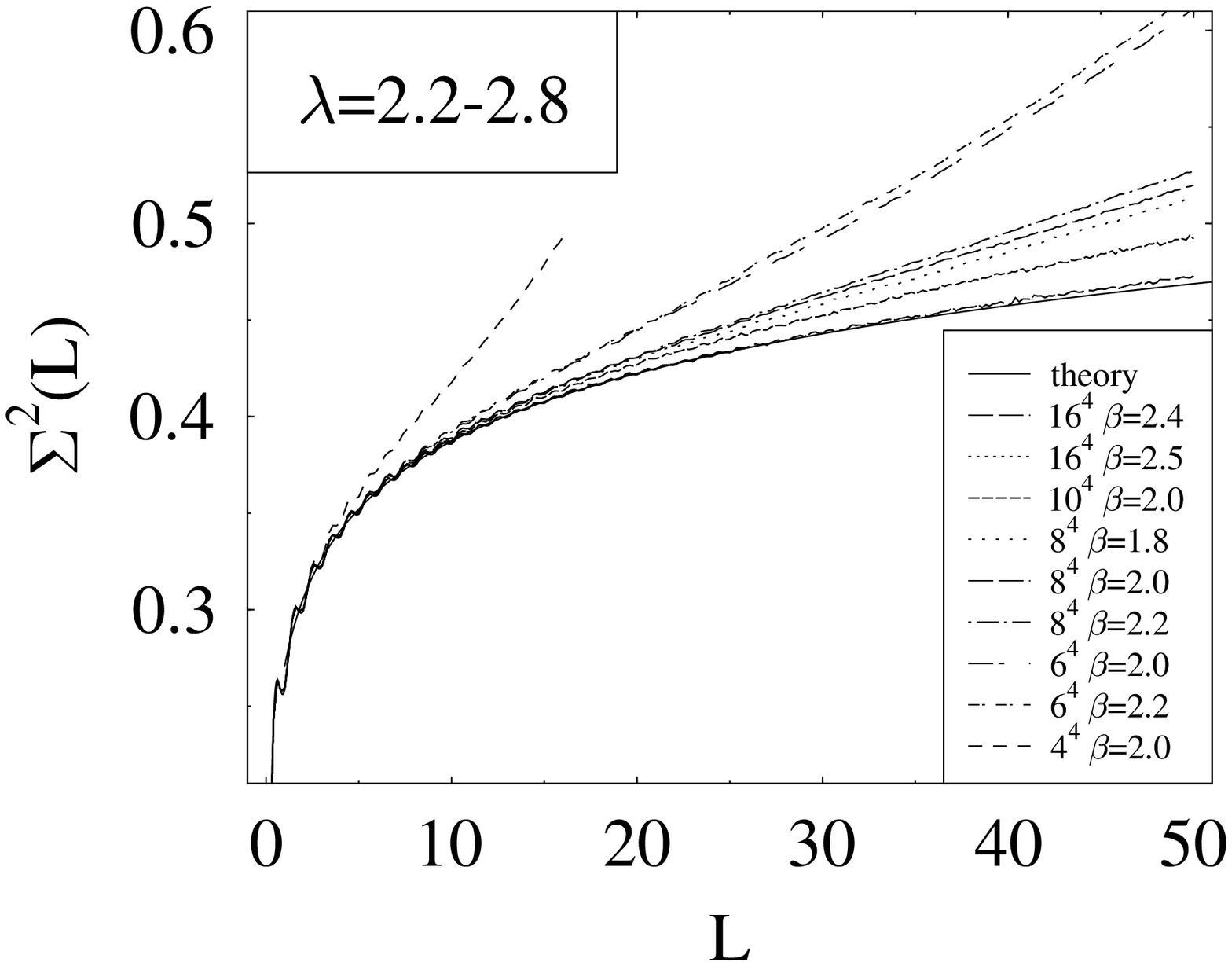,width=8cm,angle=0}}
\end{minipage}

\begin{minipage}{7.5cm}
\centerline{\psfig{figure=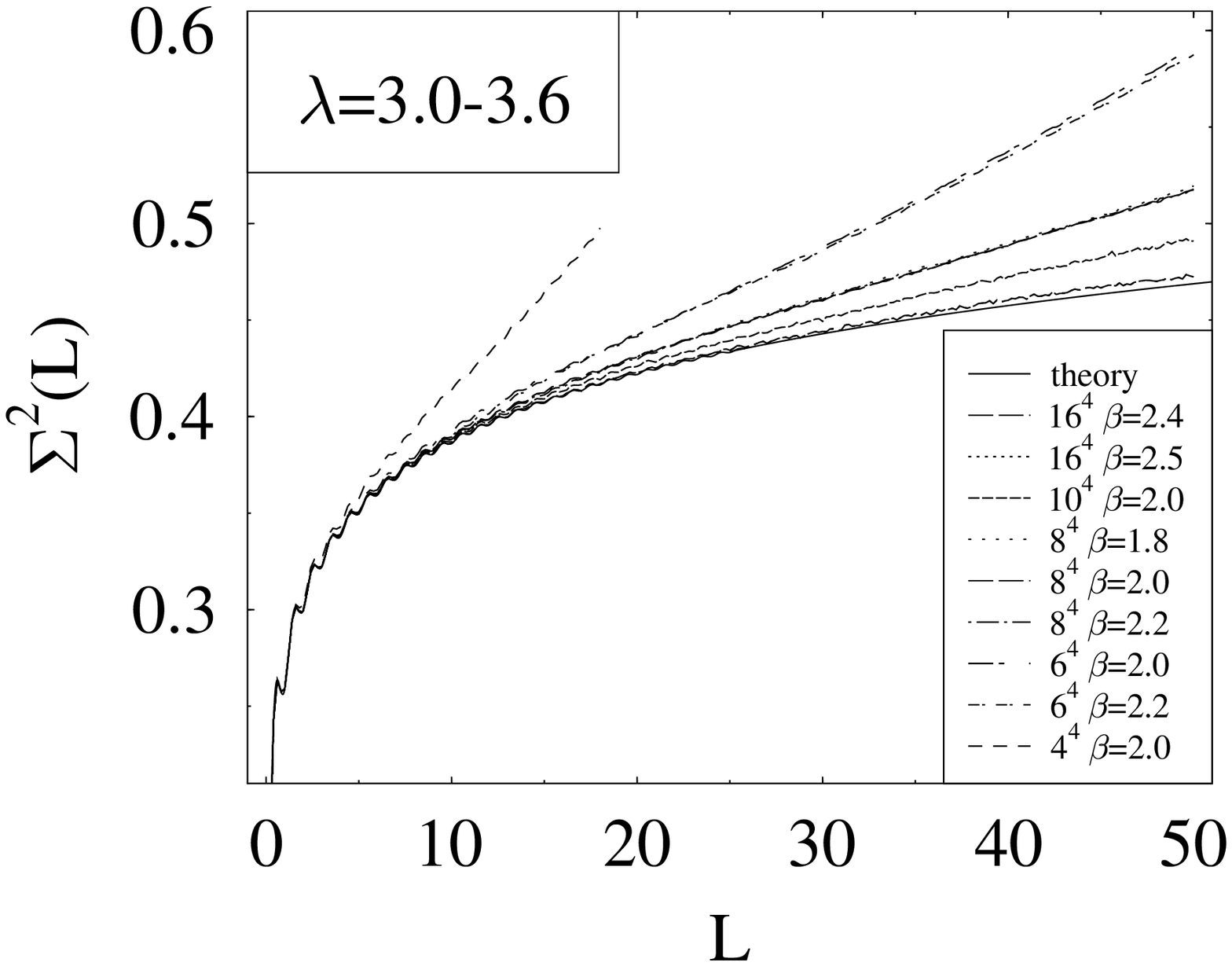,width=8cm,angle=0}}
\end{minipage}
\hfill
\begin{minipage}{7.5cm}
\centerline{\psfig{figure=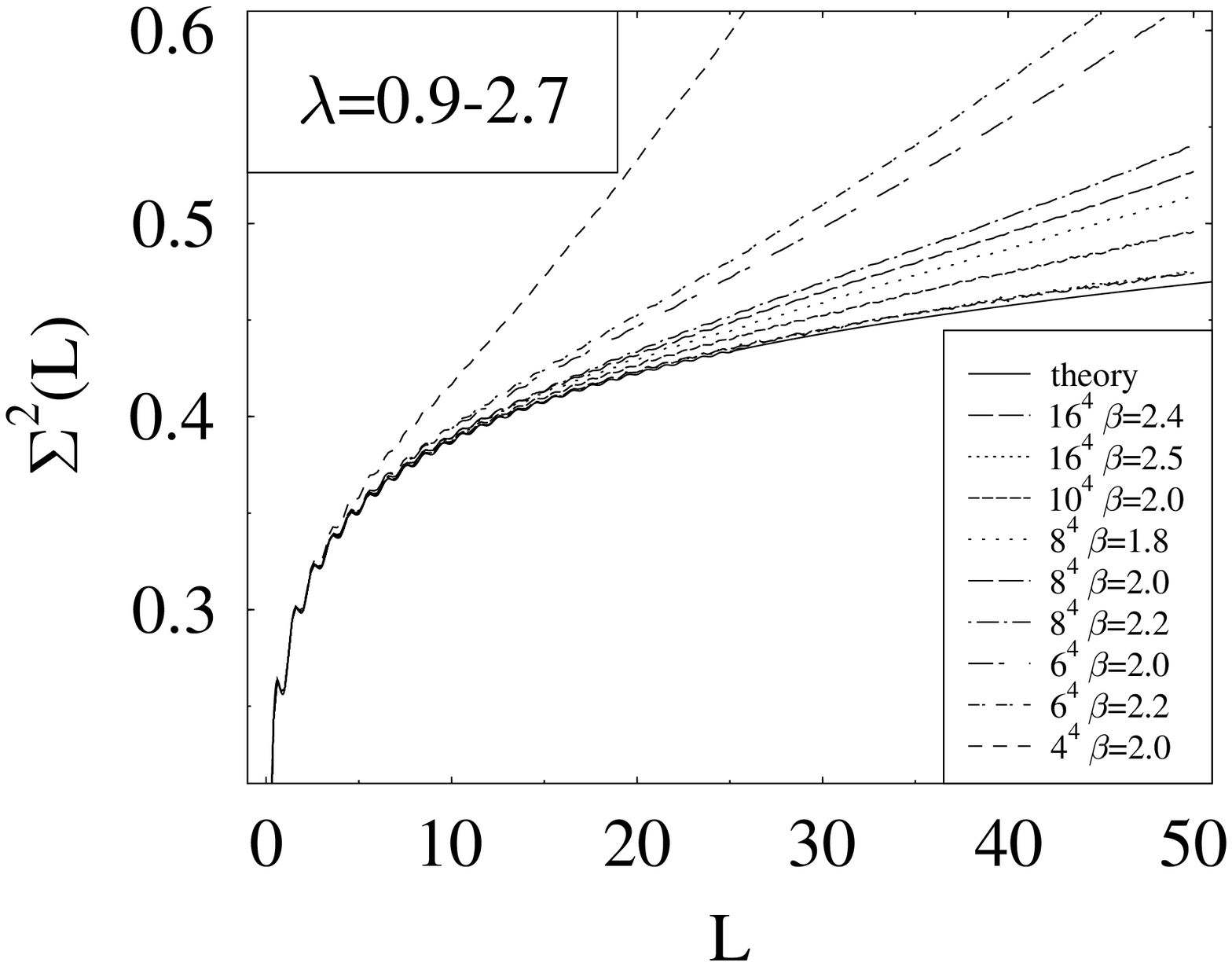,width=8cm,angle=0}}
\end{minipage}
\caption{Deviations from RMT predictions for different lattices
         sizes $V$ and gauge couplings $\beta$. Shown are different
         regions of the spectrum as indicated in the upper left
         part of the plots.
\label{enssig}}
\end{figure}

\begin{figure}
\psfig{figure=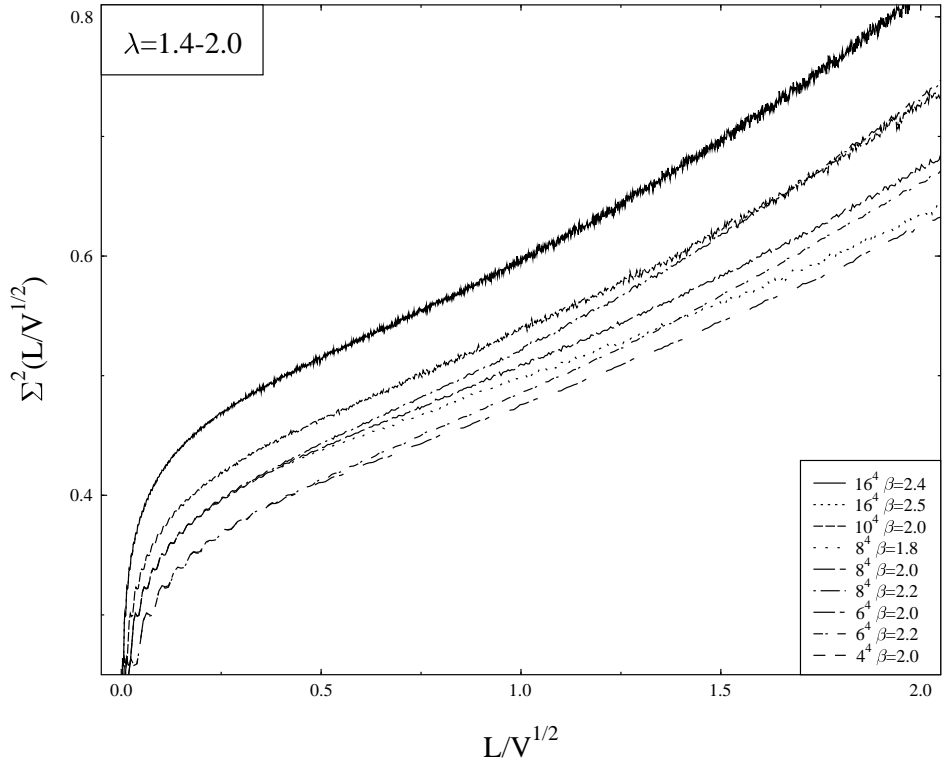,width=16cm,angle=0}
\caption{
  Deviation from the RMT predictions rescaled with the square root of
  the volume, to be compared with Fig.~\protect\ref{enssig}. The
  crossover between RMT and non--universal behavior is at
  $(\lambda_{\rm RMT}/D)V^{-1/2}\approx0.3$.
  \label{ensscl}}
\end{figure}

If we unfold with the ensemble staircase function
$\overline{N}(\lambda)$, we obtain the following results.  The number
variance can be seen in Fig.~\ref{enssig}.  Data for different lattice
sizes and different gauge couplings are shown in comparison to the RMT
predictions for some regions of the spectra. We find that the point
where the deviation sets in, scales with the square root of the
lattice volume,
\begin{equation}
\frac{\lambda_{\rm RMT}}{D}\approx C \sqrt{V}\;.
\label{scale}
\end{equation}
The numerical constant $C$ is approximately given by $C\approx0.3$.
This should be compared with the result obtained in \cite{bitsch98}
for the microscopic region. There, the scaling $\lambda_{\rm
  RMT}/D\sim0.3\ldots0.7\sqrt{V}$ was found.  This is independent of
the region of the spectra we consider and of the coupling strength
$\beta$.  This is shown in Fig.~\ref{enssig}. There different regions
of the spectra are considered, each corresponding to different values
of $\overline{\rho}(\lambda)$, see Fig.~\ref{stair}.  The results are
the same. Furthermore, the deviation points for different $\beta$
appear to coincide, whereas the local average density depends on the
gauge coupling,
$\overline{\rho}(\lambda)=\overline{\rho}(\lambda,\beta)$, see
Fig.~\ref{stair}. The scaling relation (\ref{scale}) can nicely be seen
from Fig.~\ref{ensscl}, in which the $L$ axes of Fig.~\ref{enssig} is
rescaled with $\sqrt{V}$. We see that the crossover from RMT to
non--universal behavior appears to be the same for all lattice sizes
independent of $\beta$. But the slope varies for different couplings
and regions of the spectra.  When we use windowing instead, we get the
same results as obtained by ensemble unfolding.  But the data points
scatter more compared to Figs.~\ref{enssig} and \ref{ensscl}.

\begin{figure}
\begin{minipage}{7.5cm}
\psfig{figure=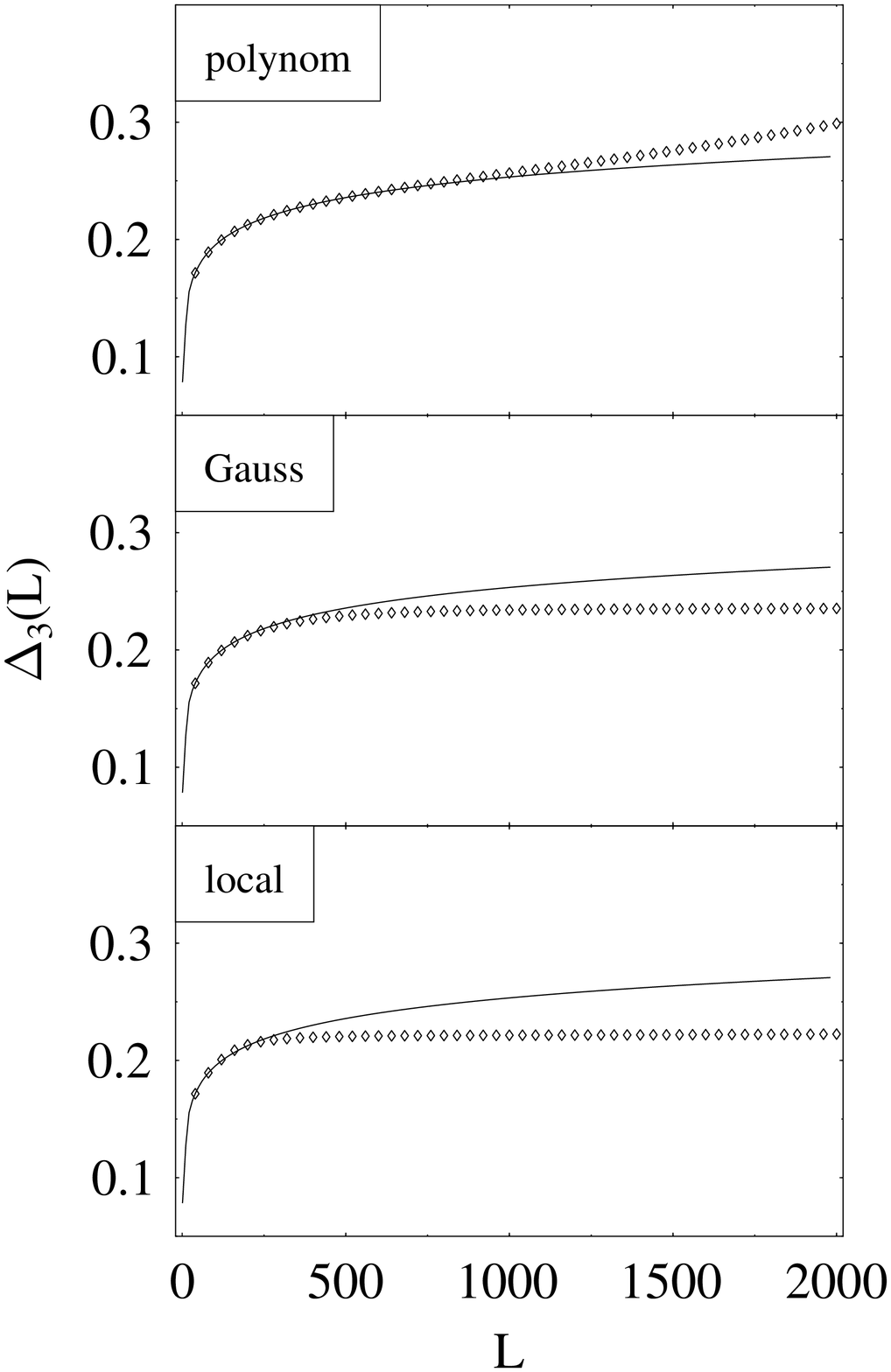,width=7.5cm,angle=0,angle=0}
\end{minipage}
\hfill
\begin{minipage}{7.5cm}
\psfig{figure=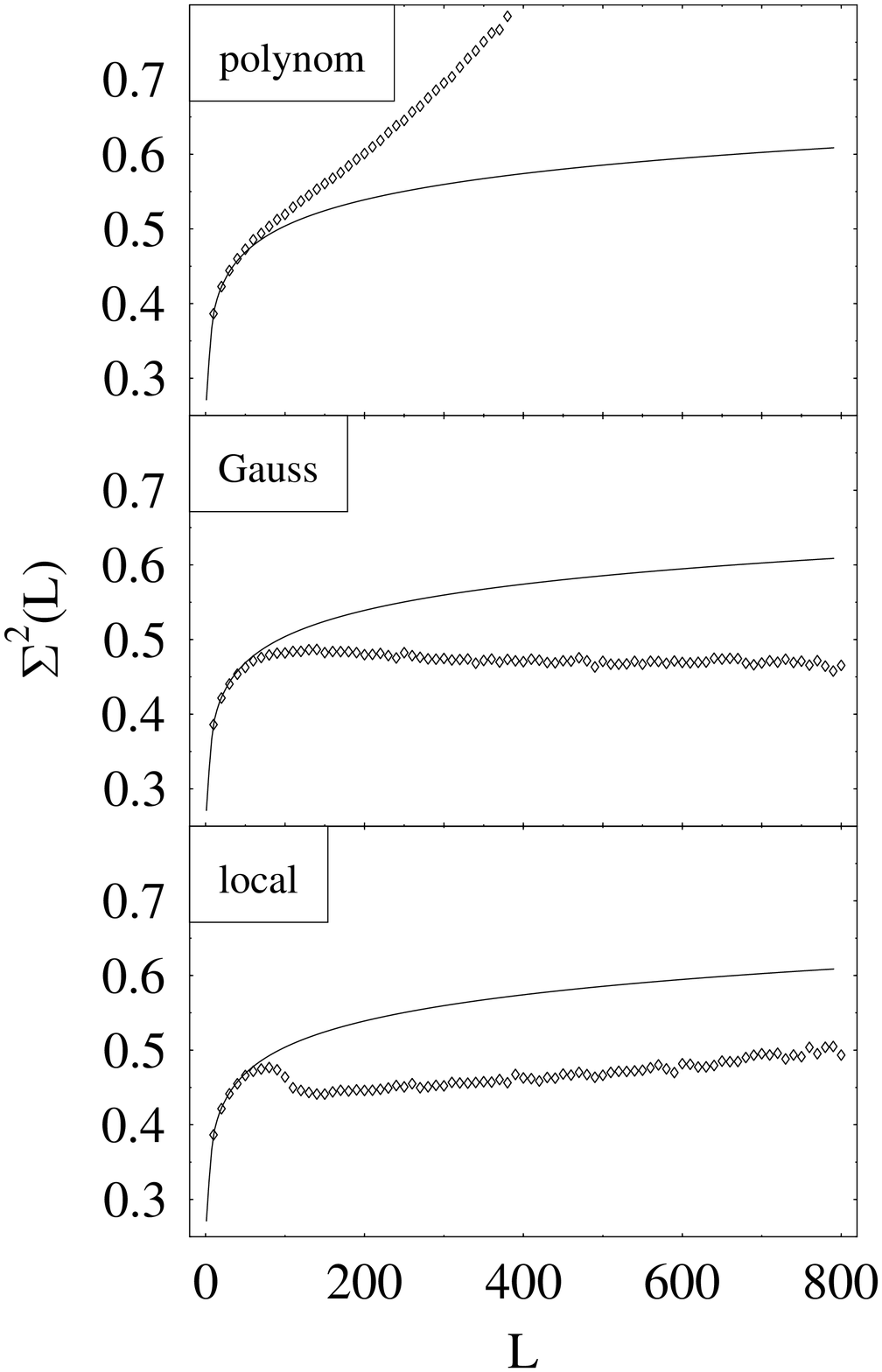,width=7.5cm,angle=0,angle=0}
\end{minipage}
\caption{Deviations of the spectral rigidity $\Delta_3(L)$ and number
  variance $\Sigma^2(L)$ from the RMT-predictions on a $16^4$-lattice
  for large $L$. From top to bottom the results for polynomial with
  degree $n=3$, Gaussian and local unfolding with averaging interval
  length $k=100$ are shown. Note the different scale on the abscissa
  compared to Fig.~\protect\ref{agree1}.
\label{deviate1}}
\end{figure}

The importance of a proper choice of the unfolding method becomes
manifest as the above picture changes drastically if we unfold each
configuration separately.  As displayed in Fig.~\ref{deviate1}, the
polynomial unfolding leads to an overshooting of the data over the
predictions but further out, compared to ensemble unfolding, while in
the Gaussian as well as in the local case $\Sigma^2(L)$ saturates.
Note the different scale compared to Fig.~\ref{enssig} and also the
difference in scale between number variance and spectral rigidity, as
mentioned above.  The result of the polynomial approach does not
depend on the degree $n$ of the polynomial. Furthermore we find no
scaling with $\sqrt{V}$ for the deviations of polynomial unfolding,
see Fig.~\ref{consig}. The deviation point appears to be the same for
different lattice sizes.  The saturation of the small lattices is due
to the limited number of eigenvalues in the considered energy range.
The same picture arises for the number variance: overshooting for the
polynomial, saturation for Gaussian and local approach.  The general
tendency of this results are already obtained for each configuration
separately, but the data points scatter. After averaging over all
configurations the scattering becomes much smaller, see
Figs.~\ref{agree1} and \ref{deviate1}.

\subsection{Higher Order Spectral Correlations}
\label{sec34}

\begin{figure}
\begin{minipage}{7.5cm}
\centerline{\psfig{figure=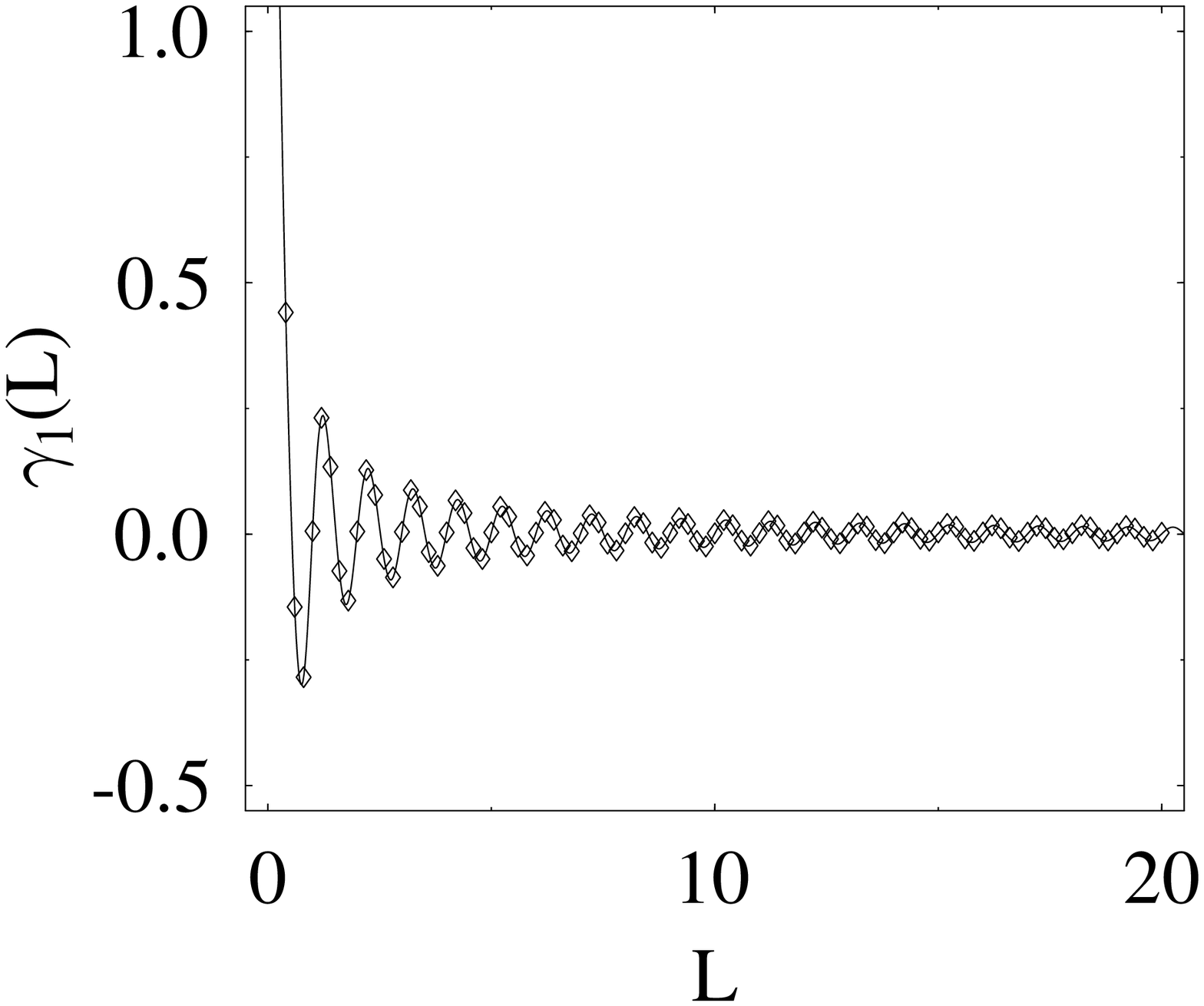,width=8cm,angle=0}}
\end{minipage}
\hfill
\begin{minipage}{7.5cm}
\centerline{\psfig{figure=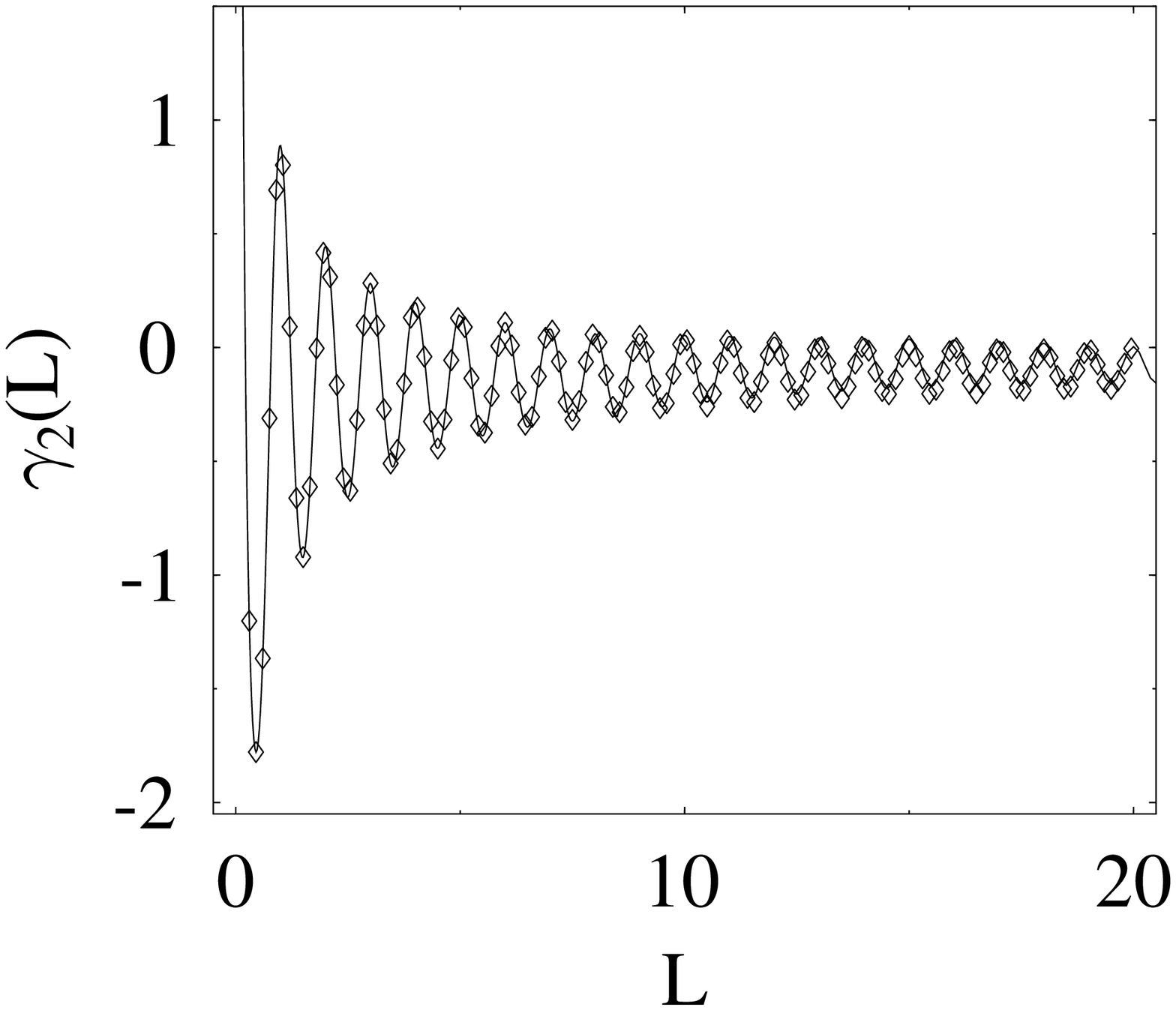,width=8cm,angle=0}}
\end{minipage}
\caption{Integrated three--point and four--point function, 
  skewness $\gamma_1(L)$ and excess $\gamma_2(L)$, respectively, as
  defined in Eqs.(\protect\ref{skew})-(\protect\ref{mu4}). The lattice
  size is $V=16^4$, as in Fig.~\protect\ref{agree1}.  The solid line
  represent the RMT predictions and the dots the data. On this scale
  the presented data points do not depend on unfolding.
  \label{agree2}}
\end{figure}

The wealth of data allows us to go beyond a previous
analysis~\cite{halasz97} of higher moments of the eigenvalue
partition,
\begin{equation}
\mu_k(L)=\overline{\langle(L-n_{\alpha}(L))^k\rangle}\;.
\label{moments}
\end{equation}
We notice that $\mu_2(L)=\Sigma^2(L)$. The skewness and the excess
\cite{Mehta} are defined by
\begin{equation}
\gamma_1(L)=\mu_3(L)\mu_2(L)^{-3/2}
\label{skew}
\end{equation}
and
\begin{equation}
\gamma_2(L)=\mu_4(L)\mu_2(L)^{-2}-3 \ ,
\label{excess}
\end{equation}
respectively.  The comparison of RMT predictions with lattice data for
these both quantities in Fig.~\ref{agree2} again shows very good
agreement.

The measures $\gamma_1(L)$ and $\gamma_2(L)$ only contain a small
amount of information of the spectral correlations. Moreover,
$\gamma_1(L)$ and $\gamma_2(L)$ also involve lower order correlations:
$\gamma_1$ is a combination of the two-- and the three--point
correlator, $Y_2(r)$ and $Y_3(r_1,r_2)$, and $\gamma_2$ involves in
addition the four--point correlator $Y_4(r_1,r_2,r_3)$. The
representation of the moments in terms of integrals over the
correlators reads
\begin{equation}
\mu_3(L) =
L-6\int_{0}^{L}(L-r)Y_2(r)dr-6\int_{0}^{L}dr_1\int_{0}^{L-r_1}dr_2
(L-r_1-r_2)Y_3(r_1,r_2)
\label{mu3}
\end{equation}
and
\begin{eqnarray}
  \mu_4(L) & = & L-(14-12L)\int_{0}^{L}(L-r)Y_2(r)dr + 12 \left[
  \int_{0}^{L}(L-r)Y_2(r)dr \right]^{2} \nonumber \\ & + & 36
  \int_{0}^{L}dr_1\int_{0}^{L-r_1}dr_2(L-r_1-r_2)Y_3(r_1,r_2)
  \nonumber \\ & - &
  24\int_{0}^{L}dr_1\int_{0}^{L-r_1}dr_2\int_0^{L-r_1-r_2}
  dr_3(L-r_1-r_2-r_3)Y_4(r_1,r_2,r_3)\;.  \nonumber\\
\label{mu4}
\end{eqnarray}
Obviously, by analyzing $\gamma_1(L)$ and $\gamma_2(L)$, one cannot
easily estimate to what extent the three-- and the four--point
correlators themselves obtained from the lattice calculations follow
the predictions of RMT.

\begin{figure}
\begin{minipage}{7.5cm}
\centerline{\psfig{figure=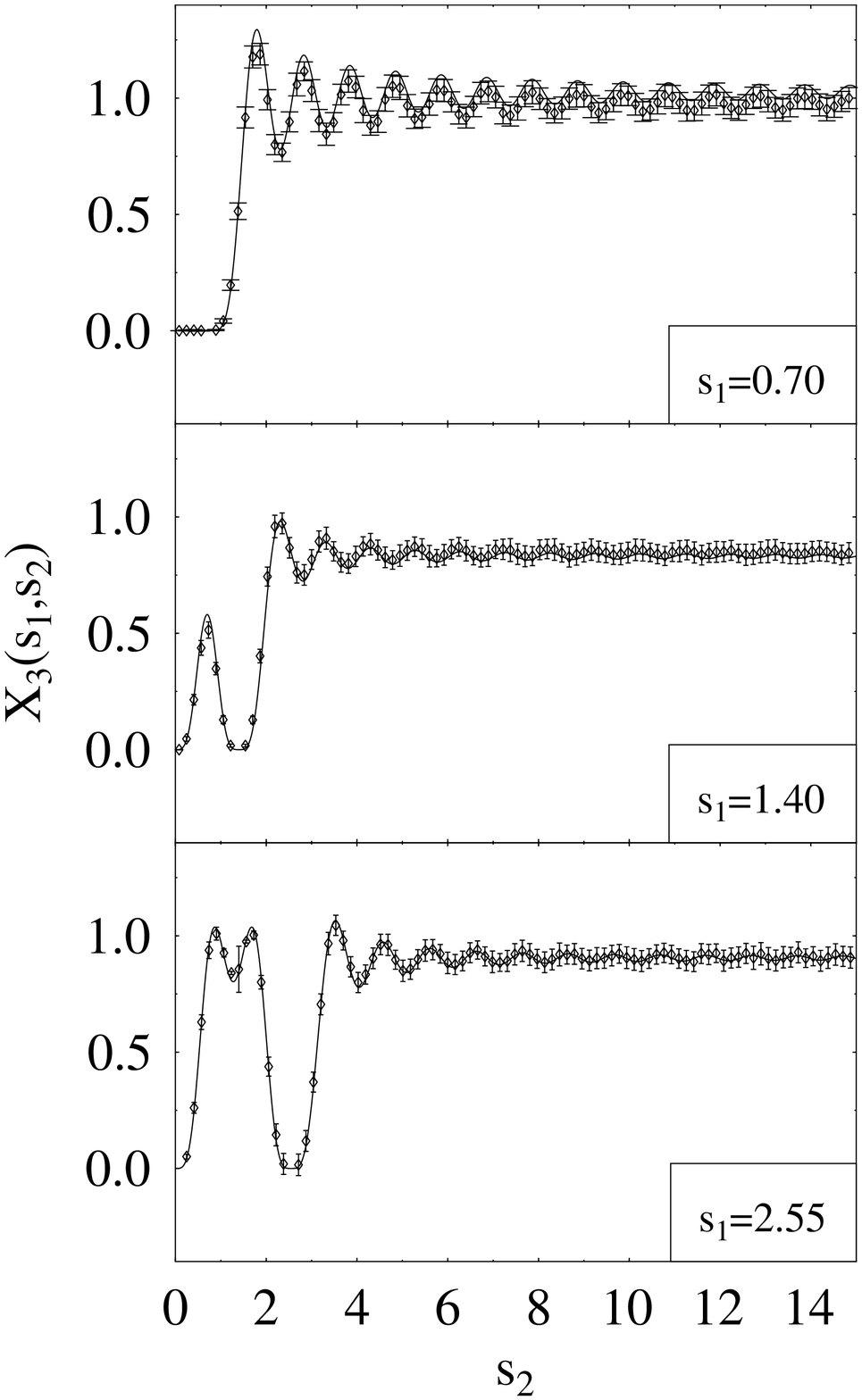,width=7.5cm,angle=0
}}
\end{minipage}
\begin{minipage}{7.5cm}
\centerline{\psfig{figure=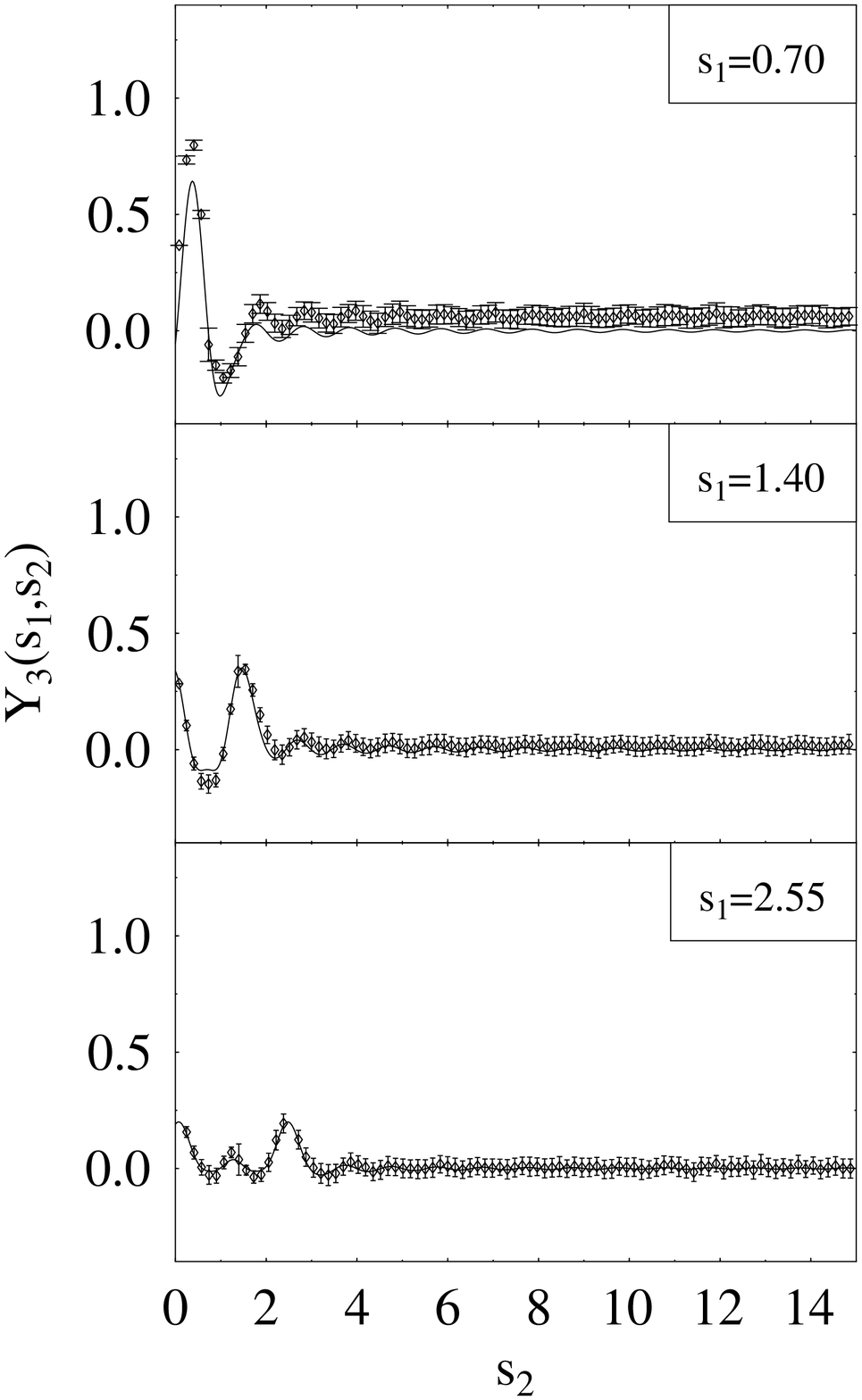,width=7.5cm,angle=0
}}
\end{minipage}
\caption{\label{twfig2}
  The three--point correlation function $X_3(s_1,s_2)$(left) and the
  cluster function $Y_3(s_1,s_2)$(right) as a function of $s_2$ for
  different values of $s_1=0.70, 1.40$ and $2.55$ (from top to
  bottom), compared with the GSE predictions. As in
  Fig.~\protect\ref{twfig1} the results are independent of the
  unfolding approach.}
\end{figure}

Here, the three-- and the four--point cluster functions,
$Y_3(r_1,r_2)$ and $Y_4(r_1,r_2,r_3)$ are written as functions of the
arguments $r_i$ ($i=1,2,3$) which are defined terms of the original
arguments $\xi_i$ ($i=1,2,3,4$) by
\begin{equation}
  r_1=\xi_2-\xi_1, \ \ r_2=\xi_3-\xi_2, \ \ r_3=\xi_4-\xi_3.
\label{arg1}
\end{equation}
We constructed from the data the two--point and the three--point
correlation functions $X_2(s)$ and $X_3(s_1,s_2)$ and the
corresponding cluster functions $Y_2(s)$ and $Y_3(s_1,s_2)$. Here, for
convenience, we redefined the arguments for the three--point
correlators as follows:
\begin{equation}
s_1=\xi_2-\xi_1=r_1, \ \ \ s_2=\xi_3-\xi_1=r_1+r_2.
\end{equation}
The results for $X_3(s_1,s_2)$ and $Y_3(s_1,s_2)$ are plotted with
error bars in Fig.~\ref{twfig2} for some given values of $s_1$.  The
results do not depend on unfolding.  In the construction of these
correlators, we first performed a spectral average by using the
translational invariance due to unfolding, and then averaged over the
ensemble. The errors for $X_3(s_1,s_2)$ and $Y_3(s_1,s_2)$ were
estimated as the variance of the ensemble fluctuations.  Once more,
very good agreement with the RMT predictions for $X_3(s_1,s_2)$ is
found, apart from a small systematic deviation which we believe can be
understood as follows. {}From the relation between $X_3(s_1,s_2)$ and
$X_2(s)$
\begin{equation}
X_3(s_1,s_2)=Y_3(s_1,s_2)-2+X_2(s_1)+X_2(s_2)+X_2(s_1-s_2),
\end{equation}
one has
\begin{equation}
X_3(s_1,s_2)|_{s_2\to \infty} = X_2(s_1).
\end{equation}
Therefore, even a small point--deviation of the two--point correlator
at $s_1$ from the theoretical predictions can result in a systematic
deviation of the whole curve $X_3(s_1,s_2)$ versus $s_2$ for this
given $s_1$.  For $Y_3(s_1,s_2)$, the quality of the agreement with
RMT is only slightly reduced, but still remarkable.  In addition to
the systematic deviation, one can see some random fluctuations around
the theoretical curves.  This is because $Y_3(s_1,s_2)$ is the
disconnected correlator, and it should represent the true three--point
correlation, while $X_3(s_1,s_2)$ also contains the two-- and
one--point functions. They play a dominate role and are in good
agreement with the corresponding GSE predictions. We notice that this
analysis was only possible due to the extremely higher number of
levels available.

\subsection{Qualitative discussion of the deviations for
  configuration unfolding}
\label{sec35}

Using ensemble unfolding, we find deviations from RMT behavior, which
scale with the square root of the volume according to the theoretical
predictions. But as we unfold each configuration separately this
effect vanishes. There are still deviations left but none of them show
a $\sqrt{V}$ scaling law.  Moreover, we have a dependence of the
results on the unfolding approach.

Concerning local and Gaussian unfolding, an explanation seems to be
easy at hand. As mentioned above, both procedures have an intrinsic,
artificial, scale. In units of the mean level spacing it has the value
$L\approx100$ for $V=16^4$ in both cases. This is approximately where
the saturation of the statistics seen in Fig.~\ref{deviate1} sets in.
We conclude that this artificial scale causes $\Sigma^2(L)$ and
$\Delta_3(L)$ to saturate. In other words, both approaches are not
capable to allow statements at scales $L\gtrsim100$.  Nevertheless,
both approaches do not show a behavior as shown in
Figs.~\ref{enssig},\ref{ensscl} for $L<100$.

\begin{figure}
\psfig{figure=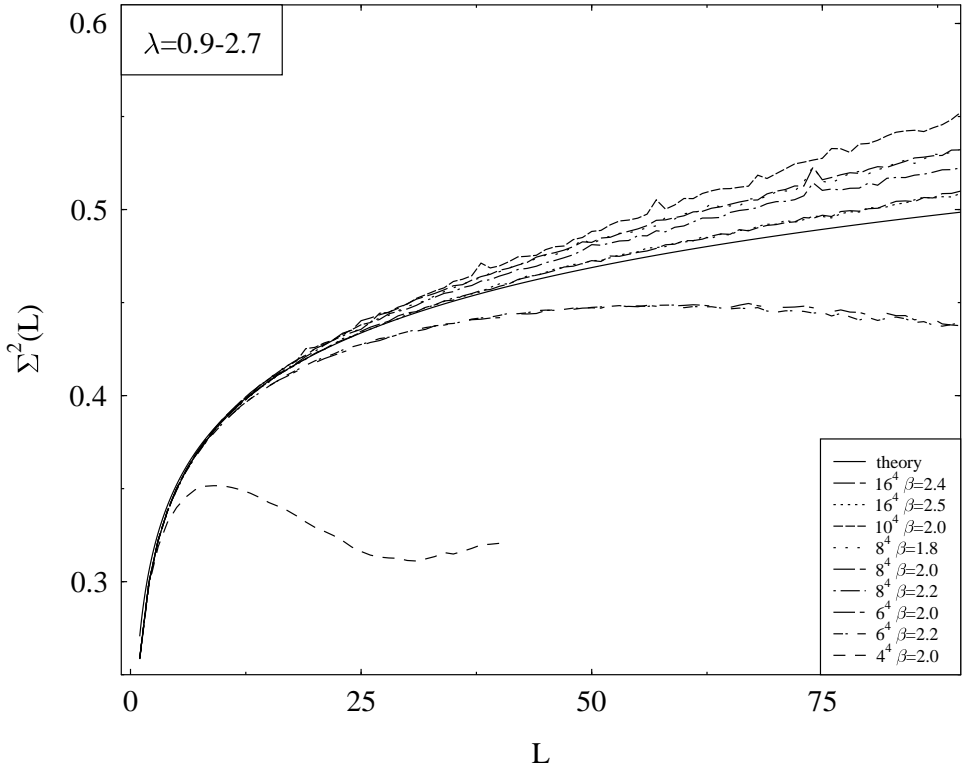,width=16cm,angle=0}
\caption{Comparison between RMT and lattice data by unfolding each
  configuration separately with a polynomial.
\label{consig}}
\end{figure}

On the other hand, the polynomial unfolding also deviates from RMT
predictions, see Fig.~\ref{consig}. As this approach does not contain
an additional scale, we can rule out effects like the one discussed
above. The question is, whether the deviations in Fig.~\ref{consig} are
due to a Thouless energy or if they have another origin.

The fluctuating part of the integrated level density $N_{\rm
  fluc}(\lambda)$ should be of order ${\cal O}(1)$, as mentioned
above.  In the upper part of Fig.~\ref{diff} the difference between the
real staircase function and the smooth polynomial staircase function,
$N(\lambda)-N_{\rm poly}(\lambda)$, for one specific configuration is
plotted. This picture remains qualitatively the same whatever
configuration is chosen.  The polynomial fits have a systematic
deviation from a smooth behavior larger than ${\cal O}(1)$. The
difference between the ensemble staircase and polynomial fit
$\overline{N}(\lambda)-N_{\rm poly}(\lambda)$ is shown in the lower
part of Fig.~\ref{diff}.  A polynomial of degree $n=3$ gives the same
result as $n=4$.  To obtain a better insight in this obviously not
universal behavior, we calculate the power spectrum
\begin{equation}
  F(f)=\int^{\infty}_{-\infty} d\lambda e^{2\pi if\lambda}
  K(\lambda_1,\lambda_2,\lambda)
  \frac{d}{d\lambda}\left(N(\lambda)-N_{\rm poly}(\lambda)\right) \ .
\label{transf}
\end{equation}
By construction, the derivative in the integrand gives the
fluctuations of the level density.  The window function
$K(\lambda_1,\lambda_2,\lambda)$ has to be introduced because we only
have a finite interval of eigenvalues in the Fourier transform over
the whole real axis. It is zero outside the interval
$\lambda_1\leq\lambda\leq\lambda_2$. The choice is not unique inside
\cite{nr}.  Thus, the Fourier transform is a convolution of the
transforms of $N(\lambda)-N_{\rm poly}(\lambda)$ and
$K(\lambda_1,\lambda_2,\lambda)$.  In order to reduce the influence of
the Fourier transform of $K(\lambda_1,\lambda_2,\lambda)$ on the
results as far as possible, we use a triangle window \cite{nr}.  The
result is shown in Fig.~\ref{spectrum}. In the right part one sees the
Fourier transform of the ensemble averaged density. Only the very
first peak is left, both for polynomial of degree $n=4$ and $n=5$.
The latter is reduced in amplitude. On the left side the transform of
an arbitrary chosen configuration is plotted. The first peaks are
reduced in amplitude again for $n=5$, whereas the remaining ones are
the same for both degrees.  We conclude that only the first peak,
corresponding to the long wave part of Fig.~\ref{diff}, is common to
all configurations. All others fluctuate from configuration to
configuration.

We conclude that the average level density $\rho_{\rm ave}(\lambda)$ and
thus the average integrated level density $N_{\rm ave}(\lambda)$
consist of two parts, namely a very smooth polynomial--like part and
another, non--universal, part,
\begin{equation}
N_{\rm ave}(\lambda)=N_{\rm poly}(\lambda)+N_{\rm osc}(\lambda)\;.
\end{equation}
We stress again that the existence of a polynomial--like smooth part
is suggested by pertubative and $1/N_c$ expansions of the QCD level
density \cite{kogut83,smilga93}. It is expected to be governed by the
available phase space: for free fermions the spectral density is given
at $|\lambda|\to\infty$ by $\rho_{\rm
  ave}(\lambda)=N_cV|\lambda|^3/4\pi^2$ \cite{leutsmilga}, where $N_c$
is the number of colors. This also holds in a $1/N_c$ expansion of
the interacting theory on scales which are large compared to the
hadronic scale \cite{smilga93}. This is the region we investigated in
the spectra, as the eigenvalues are given in units of the inverse
lattice spacing as $a^{-1}\approx10~{\rm fm}^{-1}\approx2~{\rm GeV}$,
see Figs.~\ref{stair} and \ref{diff}. This is why we tried to
approximate the average level density by a function which is as smooth
as a polynomial.

The additional structure of the level density appears to have
similarities to oscillations, see Figs.~\ref{diff} and \ref{spectrum}.
Thus, we refer to it as ``oscillatory part'', $N_{\rm osc}(\lambda)$.
This oscillatory part explains the different behavior of $\Delta_3(L)$
and $\Sigma^2(L)$ for large $L$ for different unfolding methods.  The
polynomial unfolding is clearly unable to remove the oscillations and
fits only $N_{\rm poly}(\lambda)$. Thus, the oscillatory part is still
present in the unfolded spectrum.  The presence of these oscillations
leads to values for $\Delta_3(L)$ larger than predicted by RMT,
because a fit to a straight line can only be done in a less accurate
manner. In contrast to that, the Gaussian and the local unfolding is
capable to fit $N_{\rm poly}(\lambda)$ and part of $N_{\rm
  osc}(\lambda)$.  However, it is not clear whether the fit to the
oscillatory part is done completely or if, on the other hand, it does
not smooth out part of the universal fluctuations, i.e.  overfits the
data points. But, because of the saturation of $\Sigma^2(L)$ and
$\Delta_3(L)$, see Fig.~\ref{deviate1}, we think that probably the
latter happens.

\begin{figure}
\begin{minipage}{7.5cm}
\centerline{\psfig{figure=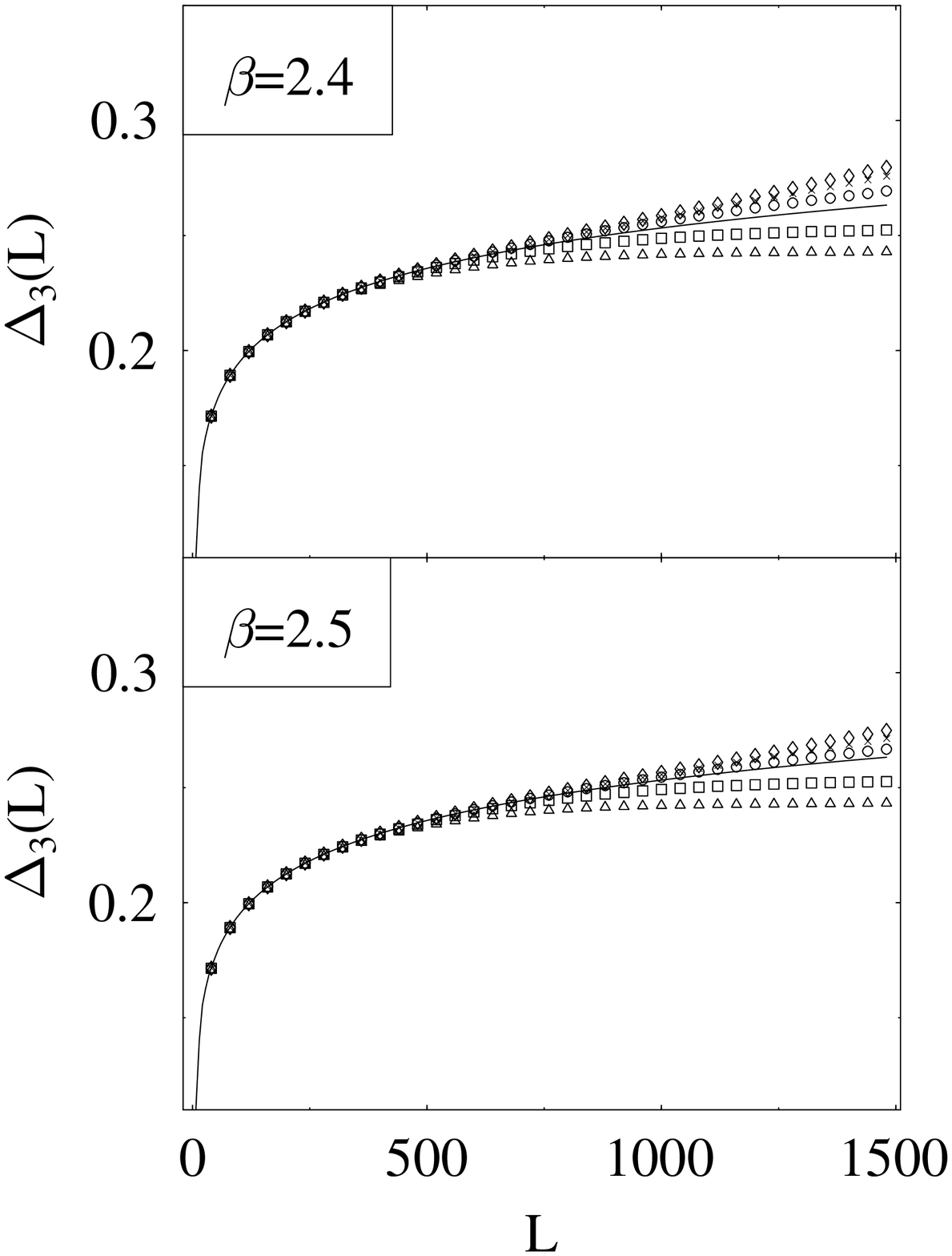,width=8cm,angle=0}}
\end{minipage}
\hfill
\begin{minipage}{7.5cm}
\centerline{\psfig{figure=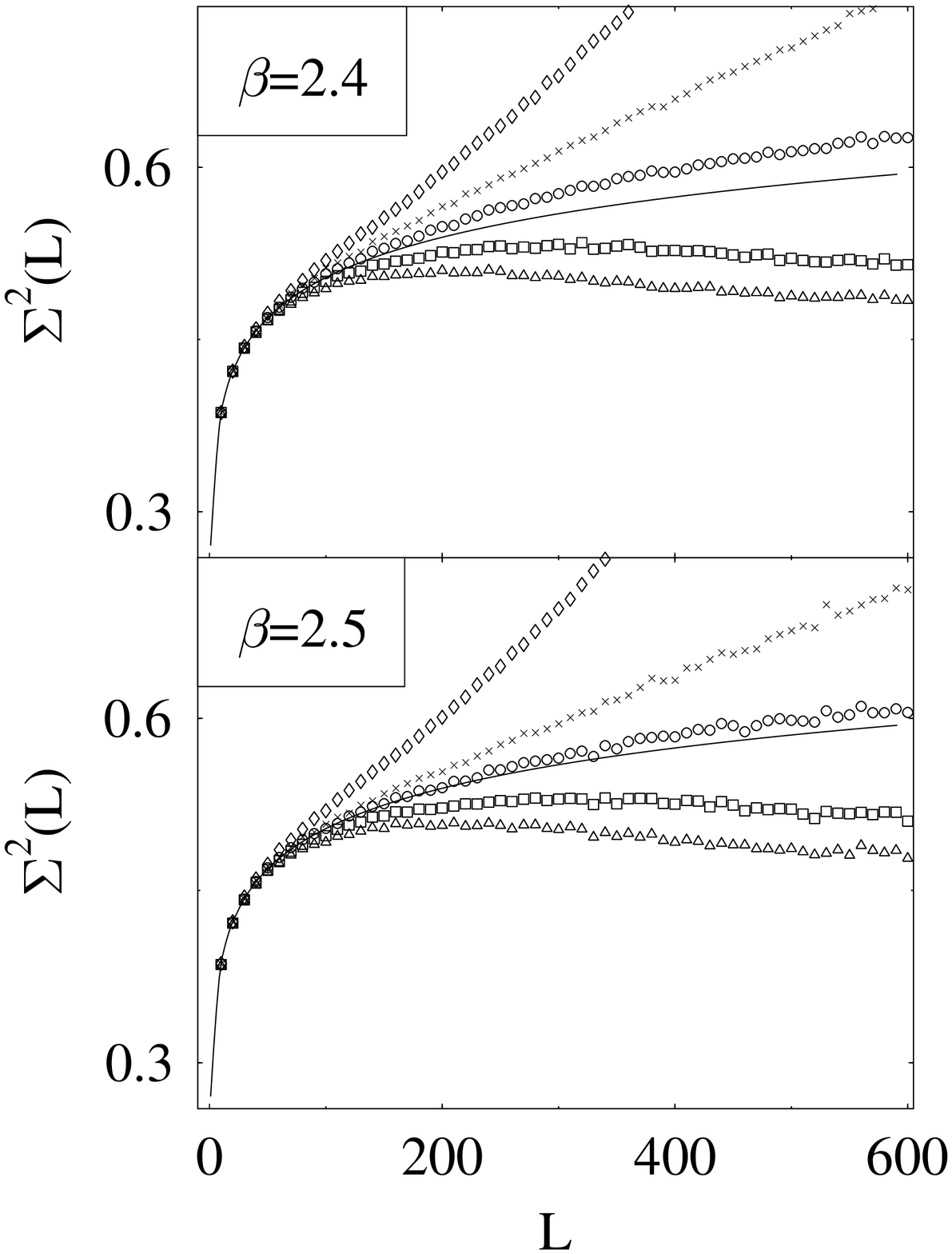,width=8cm,angle=0}}
\end{minipage}
\caption{Number variance $\Sigma^2(L)$ and
  spectral rigidity $\Delta_3(L)$ for polynomial unfolding with the
  withdraw of the long wave length oscillations, as explained in the
  text. In each plot the data corresponds from top to bottom to a cut
  of $f_{\rm cut}=0$, $1.5\cdot(2a)$, $3.0\cdot(2a)$, $7.0\cdot(2a)$
  and $10.0\cdot(2a)$, respectively.
  \label{poly_cut}}
\end{figure}

However, as argued above, since we expect the physical density to be
as smooth as a polynomial, the oscillatory part is likely to be a
lattice artifact. This is suggested by Figs.~\ref{diff} and
\ref{spectrum} which show that these oscillations live on the scale of
the inverse lattice spacing $1/a$. As $|\lambda|$ cannot be
arbitrarily large in lattice gauge theory, due to an ultraviolet
cutoff in momentum for finite lattice spacing $a$, the increase of the
density is disturbed by lattice artifacts.  For the large lattices,
i.e.  $V=16^4$, the deviations due to lattice artifacts set in at
approximately the same scale as the expected equivalent of a Thouless
energy. This can be seen from the $\sqrt{V}$ scaling of the smaller
lattices. A rough estimate gives $\lambda_{\rm RMT}/D\approx30$ for
$V=16^4$ for $\Sigma^2(L)$.  Therefore one should be careful with the
determination of the deviation point for large lattices in the bulk of
the spectra.  After removing this part from the data, we find for
$\Sigma^2(L)$ reasonable agreement with the RMT prediction up to at
least $L\approx300$ and with some uncertainties even to $L\approx500$,
see Fig.~\ref{poly_cut}.  Because of the complicated structure of the
average level density it is not possible to make any statement beyond
this scale.  In chaotic billiards, so--called bouncing ball modes
generate effects which are similar to the ones here \cite{graef92}.
In that case, however, an analytical prediction for the oscillatory
behavior was at hand. Such a result is also highly desirable in our
case.  It would be needed to furnish our phenomenological removal of
the oscillations with a theoretical justification.

In any case, we may use the information displayed in
Fig.~\ref{spectrum} to phenomenologically remove the oscillatory part.
We cut the power spectrum at a certain interval, back transform the
remaining peaks in the frequency interval $[0,f_{\rm cut}]$ and
subtract the smooth oscillatory part of the integrated level density
obtained in this manner.  We do this for each configuration
separately.  This procedure changes the result of the statistical
analysis in a crucial way, as can be seen from Fig.~\ref{poly_cut}.
There, the results for spectral measures are shown for different
choices of $f_{\rm cut}$.  Because only the very first peak is common
to all spectra whereas higher frequencies appear to be specific for
each configuration, a choice of $f_{\rm cut}>3.0\cdot(2a)$ is actually
too large. This manifests in a saturation of the statistics in
Fig.~\ref{spectrum} for $f_{\rm cut}=7.0\cdot(2a),10.0\cdot(2a)$.  A
cutoff of $f_{\rm cut}=1.5\ldots3.0\cdot(2a)$ seems to be the best
choice, but we are not able to give an exact value.  This figure also
shows a comparison between $\beta=2.4$ and $\beta=2.5$ for $V=16^4$.
Both values of $\beta$ give almost the same result.  The procedure
also removes the slight differences seen in Fig.~\ref{mean} for the
polynomials for $L\gtrsim200$. From all this, we conclude that the
deviations in Fig.~\ref{consig} are due to a non polynomial--like part
in the average level density and not due to an equivalent a Thouless
energy.  

A possible way to circumvent the problems encountered by lattice
artifacts is windowing as discussed in Sec.~\ref{sec313}.  However,
this works only if the size of the windows $\delta\lambda$ is at least
so small that the oscillations of Fig.~\ref{diff} are not important,
i.e. $\delta\lambda\ll a^{-1}$. Another way might be ensemble
unfolding, Sec.~\ref{sec311}. As the lattice artifacts are seen in each
configuration as well as in the average integrated level density this
approach might remove the oscillations. But it is by no means clear if
it really does, especially for large interval lengths.  As both
approaches give similar results for small lattices, $V\leq10^4$, for
intervals $L\lesssim20$, we conclude that the artifacts are not
important on these scales. This is supported by the observations that
deviations due to them set in at $L\approx25$ for polynomial
unfolding, independently of the lattice size, see Fig.~\ref{consig}.
Thus, these artifacts become important for analysis of larger
lattices, i.e. $V=16^4$, and for even larger ones in future
examinations.

From all this we conclude that we do not see an equivalent of a
Thouless energy if we unfold each configuration separately.  This is
in complete contrast to the results gained by ensemble unfolding. The
former surely removes the fluctuations in Fig.~\ref{ensemble}, whereas
the latter does not. We conclude that this fluctuation with respect to
the ensemble already contains the information needed to determine the
Thouless energy. After removing this information it seems that there
is no further information in the spectra, i.e. we find agreement with
RMT on huge scales.

\section{Summary and discussion}
\label{sec4}

After summarizing our results in Sec.~\ref{sec41}, we discuss our
findings in Sec.~\ref{sec42}.

\subsection{Summary}
\label{sec41}

We presented a detailed analysis of statistical properties of complete
eigenvalue spectra for staggered fermions for SU(2) lattice gauge
theory for various couplings and lattice volumes. Unfolding the data
posed certain difficulties which are not encountered in other systems.
The staircase function found by an average over the ensemble differs
in most cases from the smooth staircase of one specific configuration.
The deviations are as large as $\langle
N(\lambda)\rangle-\overline{N}(\lambda)={\cal O}(10^1)$ for $V=16^4$.
Varying the unfolding approach leads to different results for large
scales. In particular, there is a drastic difference between ensemble
and configuration unfolding.

Using ensemble unfolding, we find a range of validity of RMT, giving
$\lambda_{\rm RMT}/D=C\sqrt{V}$.  The numerical constant is
approximately given as $C\approx0.3$.  which is compatible with the
result obtained in \cite{bitsch98} for the microscopic region where
the scaling $\lambda_{\rm RMT}/D\sim0.3\ldots0.7\sqrt{V}$ was found.
The same results are obtained if we use windowing, but with less
statistical significance.

By unfolding each configuration separately, we do not see any scaling
of this type. This procedure obviously removes the fluctuations of the
staircase function relative to the ensemble average seen in
Fig.~\ref{ensemble}. We notice that fluctuations over the ensemble can
already be observed in the integrated level density. But there are
still deviations depending on the unfolding approach used.  In
particular there is an overshoot for polynomial unfolding at large
$L$. The reasons for this lies in special features of the average
level density, which consists of at least two parts: a smooth
polynomial--like and an oscillatory part.  Hence the Thouless energy
is due to the fluctuations in the ensemble.

We want to clarify the difference in the notion of ergodicity used in
spectral analysis and in lattice calculations, respectively.
Concerning the analysis of spectra, a system is called ergodic if
spectral and ensemble average yield the same results. It can be proven
rigorously that random matrices have this property in the limit of
large matrix dimension. However, here we face a different situation.
We emphasize that these issues do not affect the notion of ergodicity
as used in connection with lattice simulations. The latter is defined
by the requirement that the space of possible lattice configuration is
explored reasonably fast by the simulation algorithm.  We emphasize
that the time history of the spectra is not of importance for the
calculation of the spectral observable of Sec.~\ref{sec3}, because it
is only relevant that the configurations used, see Table \ref{tab1}, are
representative members of the ensemble, i.e.~ergodicity in the sense
of lattice simulations is fulfilled. This is ensured by our algorithms
as described in Sec.~\ref{sec2}. Furthermore, this is supported by the
observation that our results do not change if we take arbitrary
subsets of configurations, only the scattering of the data points
increases.

Furthermore we analyzed the nearest neighbor--spacing distribution,
skewness and excess for small $L$. Due to the large amount of data,
our results achieved a quality never reached before.  Due to
translational invariance in the bulk of the spectra the general
tendency for all of the observables can already be seen for one
specific configuration.  Averaging over all configurations increases
the statistical significance of the results.  The results for small
values $L$ considerably improve the statistical significance of
previous analyses\cite{halasz95,halasz97}.  To the best of our
knowledge, we presented the first statistically highly significant
analysis of bare two-- and, importantly, three--point spectral
correlators. Very good agreement between lattice data and RMT
predictions is found.  We find minor deviations for the three--point
cluster functions. In our opinion, their origin are probably very
small point--deviations of the two--point functions.

\subsection{Discussion}
\label{sec42}

\begin{figure}
\centerline{\psfig{figure=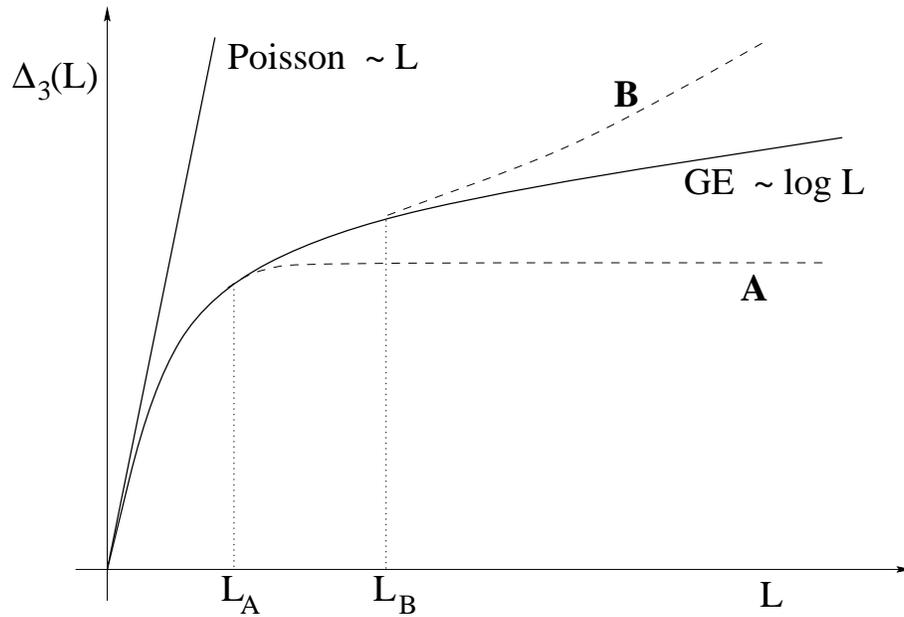,width=12cm,angle=270}}
\caption{Possible scenarios for the spectral rigidity. Plotted are
  the linear Poisson behavior and the logarithmic increase of the
  Gaussian ensembles (GE) as predicted by RMT, respectively. 'A'
  correspond to a saturation due to shortest periodic orbits and 'B'
  to linear increase due the scale set by a Thouless energy.
  \label{cartoon}}
\end{figure}

Some scenarios for the level statistics are shown in
Fig.~\ref{cartoon}.  There the spectral rigidity is plotted. The two
solid lines represent Poisson and RMT behavior, $\Delta_3(L)\sim L$
and $\Delta_3(L)\sim\log L$, respectively. The dashed lines represent
schematically two different cases.  Case `A' correspond to a scenario
known from systems with few degrees of freedom. The shortest periodic
orbit in phase space sets a scale which forces $\Delta_3(L)$ to
saturate at a certain $L_A$ \cite{berry85}. But by increasing the
degrees of freedom of the system, this scale become ever larger and is
hard to be seen.  As we may view lattice gauge theory as a many body
problem or disordered system, we do not expect to see this scale in
the level statistics.  Therefore one should not be tempted to conclude
that the saturation seen in Figs.~\ref{deviate1},\ref{poly_cut} is the
analogue of ``saturation of fluctuation measures'' which was observed
in the energy level statistics for classically chaotic systems and was
interpreted by Berry in terms of shortest periodic orbits
\cite{berry85}.  Unlike in the case of quantum billiards where an
analytic expression for the average spectral density exists, the
$N_{\rm ave}(\lambda)$ that we obtained in our analysis was only a
numerical result. Most likely all three approaches, i.e.  Gauss, local
and modified polynomial, fit also parts of the universal fluctuations
which shows up as a saturation of the statistics $\Delta_3(L)$ and
$\Sigma^2(L)$.

Another scenario is given by case `B' which lies between Poisson and
RMT behavior. It can arise in three different physical situations
\cite{guhr97}:
\begin{enumerate}
\item Systems in few degrees of freedom between regularity and
      chaos. The spectral rigidity lies between the Poisson limit,
      which applies to regular systems and the RMT limit which applies
      to chaotic systems, provided the scale $L_A$ is much larger
      than $L_B$.
\item Disordered systems in $d$ dimensions. The time $t_c$ for the
      classical diffusion through the system of size $L_S^d$ where
      $L_S$ is the size in each dimension determines an energy scale
      \begin{equation}
      E_c = \hbar / t_c \ ,
      \label{thoul}
      \end{equation}
      the Thouless energy. This in turn sets the scale
      $\lambda_{\rm RMT}/D$ where $D$ is the single particle mean
      level spacing. For energy separations in units of $D$ smaller
      than this scale, RMT fluctuations are seen. For larger energy
      separation, deviations as sketched in Fig.~\ref{cartoon} occur.
\item Many body systems such as nuclei, molecules or complex atoms.
      In the language of condensed matter physics, these systems
      are zero--dimensional. Consider the Hamiltonian of such a system
      $H=H_0+H_1$ where $H_0$ has a certain property and $H_1$ breaks this
      property. The influence of $H_1$ on the statistics of the spectrum
      of $H_0$ is measured in terms of the spreading width
      \begin{equation}
      \Gamma = 2\pi \langle H_1^2 \rangle / D_0
      \label{spread}
      \end{equation}
      where $D_0$ is the mean level spacing of $H_0$ and $\langle H_1^2
      \rangle$ the mean square matrix element of $H_1$.  In particular,
      if $H_0$ has Poisson statistics and $H_1$ RMT statistics the
      spectral rigidity acquires the form of Fig.~\ref{cartoon} and
      $\Gamma$ sets the scale $\lambda_{\rm RMT}/D$.
\end{enumerate}

As discussed above, situation 1.~is not likely to apply. We emphasize
that the Thouless energy $E_c$ and the spreading width $\Gamma$
defined in 2.~and 3.~are closely related concepts. Actually, in his
original paper, Thouless~\cite{thou75} defined $E_c$ as a special kind
of spreading width.  The precise relation between $\lambda_{\rm
  RMT}/D$ and $E_c$ has been studied in great detail in the framework
of RMT~\cite{Frahm98}. We notice that the definition in 2. which is
most commonly used in condensed matter physics, relates $E_c$ to
dimensionality, whereas $\Gamma$ is defined in zero dimensions.  Thus,
the sheer existence of a scale $\lambda_{\rm RMT}/D$ can imply, but
does not necessarily imply that the deviation from RMT is related to a
diffusion in a $d$ dimensional disordered system.

Recent theoretical studies aim at establishing
\cite{stern98,osborn98,janik98} a link between disordered systems and
QCD. In these works, a range of validity $\lambda_{\rm RMT}$ of
applicability of RMT was considered.  Using the Gell-Mann-Oakes-Renner
relation, it should be determined by the pion decay constant $F_\pi$
and the chiral condensate $\Sigma$,
\begin{equation}
\lambda_{\rm RMT}\sim\frac{F_\pi^2}{\Sigma\sqrt{V}}\;.
\label{thou1}
\end{equation}
In the microscopic region the mean level spacing is
$D=\pi/(\Sigma V)$, so that Eq.(\ref{thou1}) can be rewritten as
\begin{equation}
\frac{\lambda_{\rm RMT}}{D}\sim\frac{F_\pi^2}{\pi}\sqrt{V}\;.
\label{thou2}
\end{equation}
The predicted scaling behavior was found very recently in the
microscopic region \cite{bitsch98}.  There a crossover from RMT to
non--universal behavior was found with the scaling Eq.(\ref{thou2}).
As argued in Ref.~\cite{osborn98}, a similar relation should also be
seen in the bulk of the spectrum. Here, however, the mean level
spacing $D$ is that of the bulk. In the present work, we have shown
that, first, the scale $\lambda_{\rm RMT}/D$ exists in the bulk, and,
second, that it shows the predicted scaling behavior.  The latter
observation is a support for the ideas that link QCD to disordered
systems and thus to a diffusion in a $d$ dimensions.  However, as
outlined above, our findings are necessary but not sufficient for this
conclusion. They do not rule out other physical mechanisms that lead
to a spreading width or Thouless energy. This underlines the strength
of our analysis: it does not depend in any way on a model for this
mechanism. It is a self--consistent statistical method to find the
scale $\lambda_{\rm RMT}/D$.

We hope that the identification of this scale $\lambda_{\rm RMT}/D$
can help to improve our understanding of certain features of QCD as it
may allow one to separate the stochastic noise of the short range
fluctuations from the true dynamics of the QCD vacuum. Most desirably,
these results could help to design effective models or to simplify the
presently used simulation algorithms in lattice gauge theories.

\section*{Acknowledgments}
It is a pleasure to thank M.E. Berbenni-Bitsch for her help in
generating the data sets.  We are grateful to C. Mejia, H.J.Pirner, A.
Sch\"afer, H.A.  Weidenm\"uller and T. Wettig for enlightening
discussions.  We thank H.L. Harney and H. Alt for supplying the data
set of Ref.~\cite{graef92} for test purposes. T.G. acknowledges support
from the Heisenberg foundation.  S.M. thanks the MPI f\"ur Kernphysik
for the hospitality and acknowledges the support by a DFG grant Me
567/5-3 .

\newpage

\begin{small}
\begin{table}[!hb]
\begin{center}
   \begin{tabular}{rrrlc}
      \hline \\[-3mm]
$\beta$&$L$ &configurations&$\lambda_{\rm min}$&$\tau_{\rm int}$\\[1mm] \hline \\[-3mm]
      1.8 &  8   & 1999   &  0.00295(3) &  0.69(7)   \\
      2.0 &  4   & 9979   &  0.0699(5)  &  1.3(1)    \\
          &  6   & 4981   &  0.0127(1)  &  0.69(5)   \\
          &  8   & 3896   &  0.00401(3) &  0.71(6)   \\
          & 10   & 1416   &  0.00164(2) &   0.7(1)   \\
      2.2 &  6   & 5542   &  0.0293(3)  &   1.7(2)   \\
          &  8   & 2979   &  0.0089(1)  &   1.2(2)   \\
      2.4 &  16  &  921   &  0.00390(9) &  1.2(3)   \\
      2.5 &  8   &  576   &  0.194(9)   &  8(3)     \\
          & 16   &  543   &  0.016(2)   &  10(4)     \\[1mm] \hline
   \end{tabular}
\end{center}
   \caption{Lattice parameters and statistical analysis of the
     complete spectra of the Dirac operator.\label{tab1} }
\end{table}
\end{small}

\begin{figure}
\caption{Difference between the fitted polynom like staircase function and
  the real staircase function for one arbitrarily chosen  configuration (upper
  part). The lower part shows the difference between the staircase
  found by ensemble averaging and a polynom fit to it. The degrees of
  the polynomials are $n=4,5$. Polynomial of degree $n=3$ gives the same
  result as $n=4$.
  The plotted interval contains approximately 16000 eigenvalues.
\label{diff}}
\end{figure}

\begin{figure}
\psfig{figure=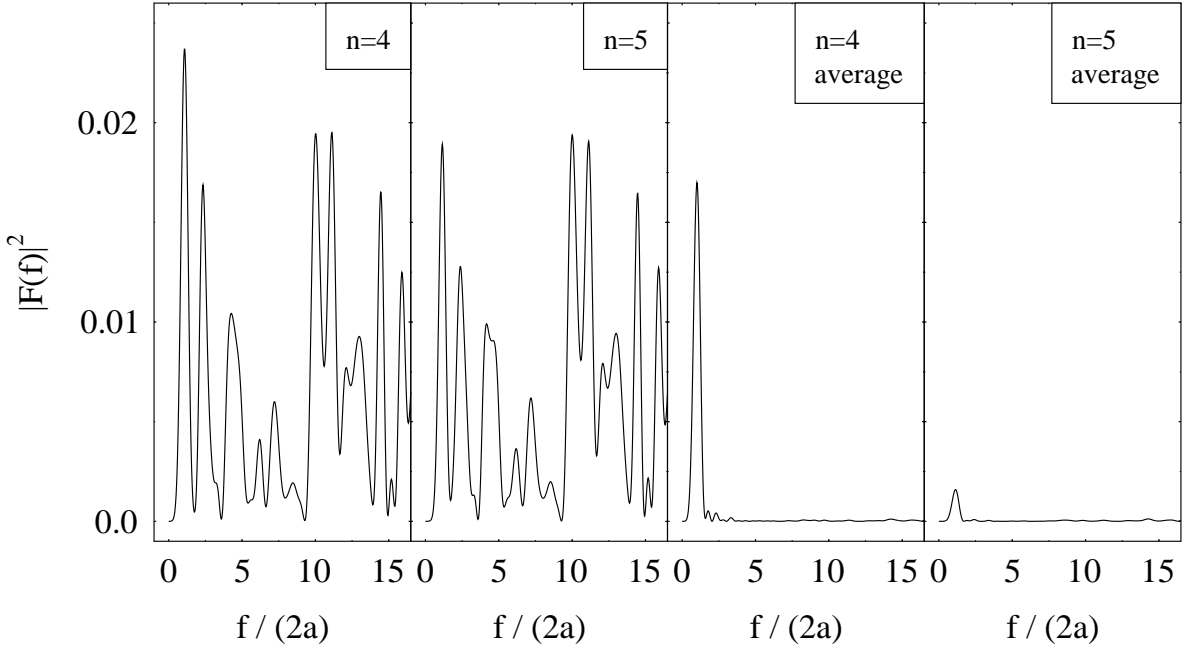,width=16cm,angle=0}
\caption{Square of the Fourier transform of the oscillation shown in
  Fig.~\protect\ref{diff}, but for
  $d(N(\lambda)-N_{\rm poly}(\lambda))/d\lambda$ instead of
  $N(\lambda)-N_{\rm poly}(\lambda)$, as given by
  Eq.(\protect\ref{transf}).
\label{spectrum}}
\end{figure}

\end{document}